\documentclass[twocolumn]{aastex63}
\usepackage{booktabs}
\usepackage{natbib}
\usepackage{multirow}
\usepackage{xcolor}

\newcommand{\msol}{\,\mbox{M$_{\odot}$}}
\shorttitle{The fast evolving type Ib Supernova SN~2015dj in NGC 7371}
\shortauthors{Singh et al.}

\begin{document}

\title{The fast evolving type Ib Supernova SN~2015dj in NGC 7371}

\correspondingauthor{Mridweeka Singh}
\email{mridweeka@kasi.re.kr, yashasvi04@gmail.com}

\author{Mridweeka Singh}
\affiliation{Aryabhatta Research Institute of observational sciencES, Manora Peak, Nainital, 263 002, India}
\affiliation{School of Studies in Physics and Astrophysics, Pandit Ravishankar Shukla University, Chattisgarh 492 010, India}
\affiliation{Korea Astronomy and Space Science Institute, 776 Daedeokdae-ro, Yuseong-gu, Daejeon 34055, Republic of Korea}

\author[0000-0003-1637-267X]{Kuntal Misra}
\affiliation{Aryabhatta Research Institute of observational sciencES, Manora Peak, Nainital, 263 002, India}
\affiliation{Department of Physics, University of California, 1 Shields Ave, Davis, CA 95616-5270, USA}

\author{Stefano Valenti}
\affiliation{Department of Physics, University of California, 1 Shields Ave, Davis, CA 95616-5270, USA}

\author{Griffin Hosseinzadeh}
\affiliation{Center for Astrophysics \textbar{} Harvard $\&$ Smithsonian, 60 Garden Street, Cambridge, MA 02138-1516, USA}

\author{Andrea Pastorello}
\affiliation{INAF Osservatorio Astronomico di Padova, Vicolo dell’Osservatorio 5, 35122 Padova, Italy}

\author{Shubham Srivastav}
\affiliation{Astrophysics Research Centre, School of Mathematics and Physics, Queen’s University Belfast, Belfast BT7 1NN, UK}

\author[0000-0002-3884-5637]{Anjasha Gangopadhyay}
\affiliation{Aryabhatta Research Institute of observational sciencES, Manora Peak, Nainital, 263 002, India}
\affiliation{School of Studies in Physics and Astrophysics, Pandit Ravishankar Shukla University, Chattisgarh 492 010, India}

\author[0000-0001-6191-7160]{Raya Dastidar}
\affiliation{Aryabhatta Research Institute of observational sciencES, Manora Peak, Nainital, 263 002, India}
\affiliation{Department of Physics $\&$ Astrophysics, University of Delhi, Delhi-110 007}

\author{Lina Tomasella}
\affiliation{INAF Osservatorio Astronomico di Padova, Vicolo dell’Osservatorio 5, 35122 Padova, Italy}

\author[0000-0001-7090-4898]{Iair Arcavi}
\affiliation{The School of Physics and Astronomy, Tel Aviv University, Tel Aviv 69978, Israel}
\affiliation{CIFAR Azrieli Global Scholars program, CIFAR, Toronto, Canada}

\author{Stefano Benetti}
\affiliation{INAF Osservatorio Astronomico di Padova, Vicolo dell’Osservatorio 5, 35122 Padova, Italy}

\author{Emma Callis}
\affiliation{School  of  Physics,  O’Brien  Centre  for  Science  North,  University College Dublin, Belfield Dublin 4, Ireland}

\author{Enrico Cappellaro}
\affiliation{INAF Osservatorio Astronomico di Padova, Vicolo dell’Osservatorio 5, 35122 Padova, Italy}

\author{Nancy Elias-Rosa}
\affiliation{INAF Osservatorio Astronomico di Padova, Vicolo dell’Osservatorio 5, 35122 Padova, Italy}
\affiliation{ Institute of Space Sciences (ICE, CSIC), Campus UAB, Carrer de Can Magrans s/n, 08193 Barcelona, Spain}

\author{D. Andrew Howell}
\affiliation{Las Cumbres Observatory, 6740 Cortona Drive Suite 102, Goleta, CA, 93117-5575 USA}
\affiliation{Department of Physics, University of California, Santa Barbara, CA 93106-9530, USA}

\author{Sang Chul Kim}
\affiliation{Korea Astronomy and Space Science Institute, 776 Daedeokdae-ro, Yuseong-gu, Daejeon 34055, Republic of Korea}
\affiliation{Korea University of Science and Technology (UST), 217 Gajeong-ro, Yuseong-gu, Daejeon 34113, Republic of Korea}

\author{Curtis McCully}
\affiliation{Las Cumbres Observatory, 6740 Cortona Drive Suite 102, Goleta, CA, 93117-5575 USA}
\affiliation{Department of Physics, University of California, Santa Barbara, CA 93106-9530, USA}

\author{Leonardo Tartaglia}
\affiliation{Department of Astronomy and The Oskar Klein Centre, AlbaNova University Centre, Stockholm University, SE-106 91 Stockholm, Sweden}

\author{Giacomo Terreran}
\affiliation{Center for Interdisciplinary Exploration and Research in Astrophysics (CIERA) and Department of Physics and Astronomy, Northwestern University, Evanston, IL 60208, USA}

\author{Massimo Turatto}
\affiliation{INAF Osservatorio Astronomico di Padova, Vicolo dell’Osservatorio 5, 35122 Padova, Italy}




\begin{abstract}

We present the detailed optical evolution of a type Ib SN 2015dj in NGC 7371, using data spanning up to $\sim$ 170 days after discovery. SN 2015dj shares similarity in light curve shape with SN 2007gr and peaks at M$_{V}$ = $-17.37\pm$0.02 mag. Analytical modelling of the quasi bolometric light curve yields 0.06$\pm$0.01 M$_{\odot}$ of $^{56}$Ni, ejecta mass $M_{\rm ej} = 1.4^{+1.3}_{-0.5}$ \msol\, and kinetic energy $E_{\rm k} = 0.7^{+0.6}_{-0.3} \times 10^{51}$ erg. The spectral features show a fast evolution and resemble those of spherically symmetric ejecta. The analysis of nebular phase spectral lines indicate a progenitor mass between 13-20 M$_{\odot}$ suggesting a binary scenario.

\end{abstract}


\section{Introduction}
\label{Introduction}

Type Ib Supernovae (SNe) have two commonly accepted progenitor scenarios - (i) massive Wolf Rayet (WR) stars that lose their hydrogen envelope by stripping via strong stellar winds \citep{1986ApJ...306L..77G} and, (ii) lower mass progenitors in close binary systems \citep{1992ApJ...391..246P,1995PhR...256..173N,2009ARA&A..47...63S}. The association of hydrogen deficient WR stars as progenitors of type Ib SNe has not been confirmed yet {\citep{2009ARA&A..47...63S}}. Previous attempts for direct identification of progenitors of type Ib SNe \citep{2007MNRAS.381..835C,2009ARA&A..47...63S,2013MNRAS.436..774E} were unsuccessful. However, in the case of iPTF13bvn, a possible progenitor identification was reported by \cite{2013ApJ...775L...7C}.
Based on observational constraints, both a single WR star and an interacting binary progenitor were proposed \citep{2013A&A...558L...1G,2014A&A...565A.114F, 2016ApJ...825L..22F, 2014AJ....148...68B,2015MNRAS.446.2689E,2015A&A...579A..95K,2016A&A...593A..68F,2016MNRAS.461L.117E,2017MNRAS.466.3775H,2017MNRAS.469L..94H}.

If both H and He layers are stripped in the progenitor star, the early time spectra are those of a type Ic SN, without H and He lines. Nonetheless, the lack of these features does not necessarily imply the absence of these two elements in the SN ejecta  \citep{2002ApJ...566.1005B, 2006PASP..118..791B, 2006A&A...450..305E,2012MNRAS.422...70H}. Hints of H and He lines, occasionally detected in type Ib/Ic SNe are due to a thin residual layer containing these elements. This indicates that there may be a continuity in the spectra of type Ib/Ic SNe. 

Late-time spectra of all stripped envelope SNe show strong emission lines of oxygen and calcium. These nebular lines are representative of the one dimensional line-of-sight projection of the three dimensional distribution of these elements \citep{2008Sci...319.1220M, 2008ApJ...687L...9M,2009MNRAS.397..677T}. The study of the nebular lines is crucial to estimate the mass of neutral oxygen and He core.

A few observational features play a key role in exploring the properties of type Ib/Ic SNe. The rise and the peak of the light curve are two of the important parameters to probe the amount of radioactive $^{56}$Ni synthesized in the explosion. This phase of light curve gives insight on the emission from the cooling envelope, which is produced when the SN ejecta radiates the energy from the SN shock \citep{2013ApJ...769...67P}. Observing type Ib/Ic SNe at early stages allows us to investigate the progenitor structure before the explosion \citep{2010ApJ...725..904N,2011ApJ...728...63R,2014ApJ...788..193N} and the degree of mixing of  radioactive material within the outer ejecta \citep{2012MNRAS.424.2139D,2013ApJ...769...67P}. 

\cite{2015MNRAS.450.1295W} discusses about the explosion parameters derived at peak and tail phase of the light curves of type Ib/Ic SNe and states that discrepancy may lie under the methods and simplified assumptions associated with the models used. The estimated physical parameters are sensitive to the way the data are selected and employed. Consistent methods have been proposed by \cite{2014MNRAS.438.2924C,2014MNRAS.437.3848L,2015A&A...574A..60T} to get the physics of the peak to agree with the physics of the tail. Heterogeneity exists in different light curves depending upon adopted methodologies, but, no correlation have been established between SN type and late time properties \citep{2018A&A...609A.136T,2019MNRAS.485.1559P}. 

This paper is organised in the following manner. Section \ref{SN 2015dj: Data and reduction} details about the data obtained and reduction procedures used. Section \ref{Explosion epoch, distance and Extinction} gives detailed information about the explosion epoch, extinction and distance adopted throughout the paper. Section \ref{Temporal evolution of SN 2015dj} provides complete analysis of the photometric properties of the SN. A comprehensive picture of the spectroscopic features is given in section \ref{spectral_evolution}. A concise summary is presented at the end of the paper in section \ref{Summary}.

\begin{figure*}[ht]
		\includegraphics[width=1.0\textwidth]{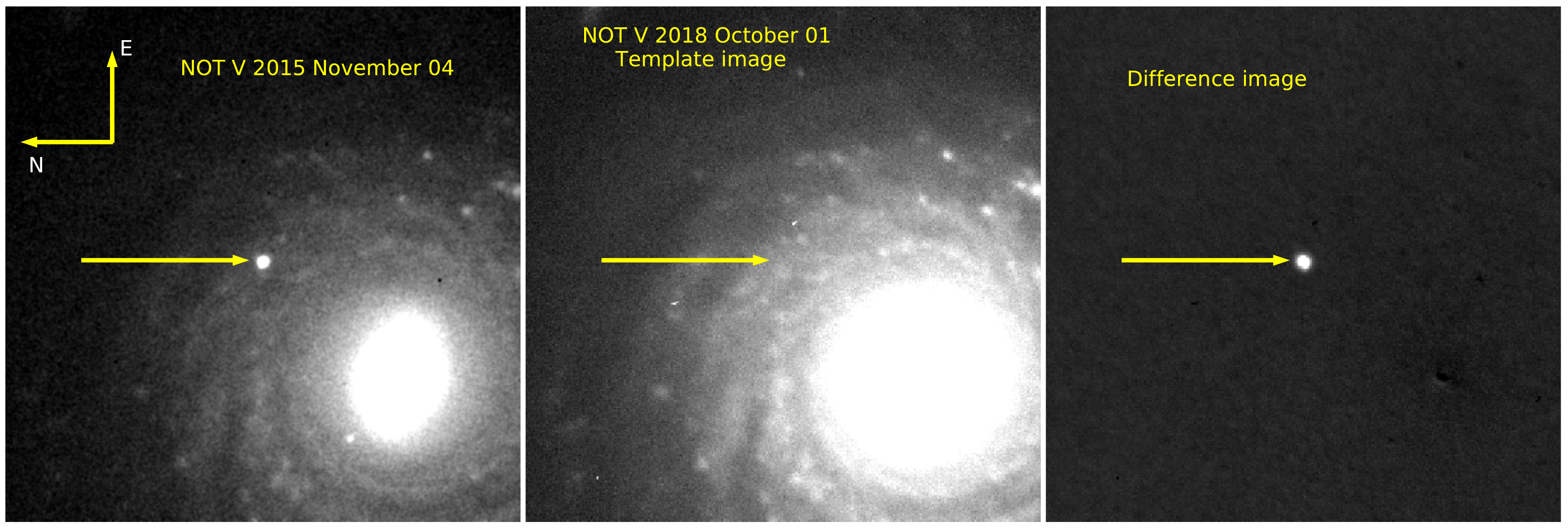}
	\caption{The left and middle panels show the image of the SN and template acquired with the 2.56 m NOT and the difference image is shown in the rightmost panel.}
	\label{fig:ds9_image}
\end{figure*}

\section{Data and reduction}
\label{SN 2015dj: Data and reduction}

SN 2015dj (R.A. 22$^h$46$^m$05.04$^s$ and Dec. $-$10$^d$59$^m$48.4$^s$) was discovered by Koichi Itagaki on 2015 July 10.655 (UT) in the galaxy NGC 7371, at an unfiltered magnitude of 16.7 mag. The SN was located 19$''$ east and 16$''$ north of the center of NGC 7371\footnote{\url{http://www.cbat.eps.harvard.edu/unconf/followups/J22460504-1059484.html}}. SN 2015dj is the IAU name for the transient and other aliases of SN 2015dj are PSN J22460504-1059484 and PS15bgt. The SN was initially classified as type IIb \citep{2015ATel.7787....1T}, but was later reclassified as a type Ib SN by \cite{2015ATel.7821....1S}.

The initial classification of the SN based on {\sc snid}  \citep{2007ApJ...666.1024B} and {\sc gelato}  \citep{2008A&A...488..383H} was inconclusive, owing to its similarity with a type IIb (SN 2000H) and several type Ib events. The absorption feature at $\sim$6160\AA{} in the classification spectrum was probably identified in {\sc snid} and {\sc gelato} as the high velocity absorption component of H$\alpha$, commonly found in the early spectra of type IIb SN. However, the absence of a discernible H$\beta$ feature in the classification spectrum suggests that the feature rather corresponds to Si II 6355\AA. Since Si II 6355\AA{} is prevalent in the early spectrum of type Ib SNe, we conclude that SN 2015dj is indeed a type Ib event.

\citet{2015ATel.7845....1K} reported a detection in the radio bands using the observations carried out with the VLA on 2015 July 23.4 UT.  Utilising the two, $\sim$1 GHz sidebands centered at 5 and 7 GHz, the flux of the detected radio source was reported to be 1.0$\pm$0.03 mJy in the upperside of 7 GHz. No further radio observations were reported in the literature.

The photometric and spectroscopic follow up of SN 2015dj initiated soon after discovery and lasted $\sim$ 170 days, mostly obtained through the Las Cumbres Observatory (LCO, \cite{2013PASP..125.1031B}) Supernova Key Project. The photometry was done with the 1m LCO telescopes and the spectroscopy was done with the FLOYDS spectrograph on the 2m LCO telescopes. Additional observations were done with the 1.82m Copernico telescope (Asiago, Italy) and the 2.56m Nordic Optical Telescope (NOT). Broadband {\it BVgriz} filters and grisms in the wavelength range of 3300--10000 \AA{} were used in the observational campaign of SN 2015dj. 

The LCO photometry was performed using the \texttt{lcogtsnpipe} pipeline  \citep{2016MNRAS.459.3939V}. Since the SN is located close to the host galaxy, template subtraction was adopted to estimate the true SN magnitude after removing the host galaxy contamination. The templates were observed on 2016 June 6 and June 18 in {\it BVgr} and {\it i} bands respectively using 1m LCO telescopes, which is approximately 1 year after the explosion. Since the SN could be detected with 1.82 m Copernico telescope on a later date (2016 August 08), we have included the upper limits on the SN magnitudes for the LCO template images in Table 3. The image subtraction was performed using PyZOGY \citep{2017zndo...1043973G} on LCO data. The 1.82 m Copernico telescope and the 2.56 m NOT data were processed within the IRAF\footnote{IRAF stands for Image Reduction and Analysis Facility distributed by the National Optical Astronomy Observatories which is operated by the Association of Universities for research in Astronomy, Inc., under cooperative agreement with the National Science Foundation.} environment and the instrumental magnitudes were derived from the point spread function photometry using DAOPHOT2 \citep{1987PASP...99..191S}. The images taken on 2018 October 01 in {\it BVgriz} bands with 2.56m NOT were used for template subtraction of the images acquired with 1.82m Copernico telescope and 2.56m NOT following the method described in \cite{2019MNRAS.485.5438S}.
 
Figure \ref{fig:ds9_image} shows the difference imaging performed in one of the images taken with the 2.56 m NOT. 
The instrumental SN magnitudes were calibrated with respect to the APASS catalog. Table \ref{tab:photometric_observational_log} gives the SN magnitudes and the associated photometric errors.

\begin{table*}
\caption{Optical photometric data}
\centering
\smallskip
\scriptsize
\begin{tabular}{c c c c c c c c c c}
\hline \hline
Date    &   JD$^\dagger$   &   Phase$^\ddagger$ &   B       &      V            &  g                & r   	         & i                   & z & Telescope$^\star$\\
        &                  &   (Days)           & (mag)     & (mag)             & (mag)             &(mag) 		&(mag)       &(mag)     \\
\hline  
 2015-07-23 & 226.80 &  -3.47 &  17.12  $\pm$   0.06  &   16.38 $\pm$ 0.01 & 16.74 $\pm$  0.02  & 16.17  $\pm$ 0.02  &  16.03 $\pm$  0.03 & & 1 \\ 
 2015-07-23 & 226.80 &  -3.47 &  17.07  $\pm$   0.03  &   16.32 $\pm$ 0.01 & 16.70 $\pm$  0.03  & 16.07  $\pm$ 0.02  &  16.02 $\pm$  0.03 & & 1 \\ 
 2015-07-26 & 230.26 &  -0.01 &  17.26  $\pm$   0.05  &   16.23 $\pm$ 0.02 & 16.52 $\pm$  0.02  & 15.96  $\pm$ 0.01  &  15.81 $\pm$  0.02 & & 1 \\ 
 2015-07-26 & 230.26 &  -0.01 &  17.19  $\pm$   0.05  &   16.21 $\pm$ 0.03 & 16.54 $\pm$  0.02  & 16.00  $\pm$ 0.01  &  15.71 $\pm$  0.03 & & 1 \\ 
 2015-07-30 & 234.04 &  3.76  &  17.62  $\pm$   0.10  &   16.39 $\pm$ 0.07 & 16.92 $\pm$  0.07  & 16.13  $\pm$ 0.04  &  16.04 $\pm$  0.13 & & 1 \\ 
 2015-07-30 & 234.04 &  3.76  &  17.56  $\pm$   0.09  &   16.33 $\pm$ 0.05 & 16.90 $\pm$  0.06  & 16.11  $\pm$ 0.04  &  15.96 $\pm$  0.06 & & 1 \\ 
 2015-08-05 & 239.84 &  9.56  &  18.38  $\pm$   0.14  &   16.95 $\pm$ 0.05 & 17.75 $\pm$  0.07  & 16.63  $\pm$ 0.04  &  16.30 $\pm$  0.04 & & 1 \\ 
 2015-08-05 & 239.85 &  9.57  &  18.39  $\pm$   0.13  &   16.91 $\pm$ 0.05 & 17.70 $\pm$  0.09  & 16.71  $\pm$ 0.04  &  16.35 $\pm$  0.03 & & 1 \\ 
 2015-08-06 & 241.52 &  11.25 &  18.53  $\pm$   0.13  &   17.02 $\pm$ 0.01 & 17.70 $\pm$  0.01  & 16.84  $\pm$ 0.02  &  16.61 $\pm$  0.02   & 16.42$\pm$0.03& 2 \\ 
 2015-08-08 & 243.44 &  13.16 &  18.88  $\pm$   0.15  &   17.31 $\pm$ 0.04 & 17.99 $\pm$  0.05  & 16.89  $\pm$ 0.04  &  16.45 $\pm$  0.03 & & 1 \\ 
 2015-08-08 & 243.45 &  13.17 &  19.15  $\pm$   0.19  &   17.16 $\pm$ 0.03 & 18.08 $\pm$  0.06  & 16.88  $\pm$ 0.03  &  16.36 $\pm$  0.03 & & 1 \\
 2015-08-10 & 245.53 &  15.25 &  18.84  $\pm$   0.11  &   17.29 $\pm$ 0.01 & 18.04 $\pm$  0.01  & 16.60  $\pm$ 0.01  &  16.78 $\pm$  0.01   & 16.69$\pm$0.06& 2 \\  
 2015-08-11 & 246.57 &  16.29 &  18.90  $\pm$   0.12  &   17.35 $\pm$ 0.02 & 18.18 $\pm$  0.01  & 16.81  $\pm$ 0.01  &  16.86 $\pm$  0.04   & 17.09$\pm$0.02& 2 \\  
 2015-08-12 & 247.35 &  17.07 &  18.58  $\pm$   0.19  &   17.38 $\pm$ 0.05 & 18.25 $\pm$  0.07  & 16.95  $\pm$ 0.03  &  16.43 $\pm$  0.04 & & 1 \\ 
 2015-08-12 & 247.35 &  17.07 &  18.87  $\pm$   0.19  &   17.41 $\pm$ 0.06 & 18.27 $\pm$  0.07  & 16.97  $\pm$ 0.03  &  16.50 $\pm$  0.03 & & 1 \\ 
 2015-08-16 & 251.15 &  20.87 &  19.48  $\pm$   0.21  &   17.60 $\pm$ 0.05 & --                 & 17.24  $\pm$ 0.04  &  16.67 $\pm$  0.04 & & 1 \\ 
 2015-08-16 & 251.16 &  20.88 &  --                   &   --               & --                 & --                 &  16.63 $\pm$  0.04 & & 1 \\
 2015-08-22 & 256.57 &  26.29 &  19.33  $\pm$   0.25  &   17.81 $\pm$ 0.08 & 18.76 $\pm$  0.10  & 17.34  $\pm$ 0.06  &  16.79 $\pm$  0.05 & & 1 \\ 
 2015-08-26 & 260.91 &  30.63 &  19.23  $\pm$   0.20  &   17.81 $\pm$ 0.07 & 18.51 $\pm$  0.08  & 17.43  $\pm$ 0.05  &  16.93 $\pm$  0.05 & & 1 \\ 
 2015-08-27 & 262.23 &  31.95 &  19.44  $\pm$   0.26  &   17.99 $\pm$ 0.09 & 18.60 $\pm$  0.13  & 17.54  $\pm$ 0.08  &  17.11 $\pm$  0.06 & & 1 \\ 
 2015-08-31 & 265.80 &  35.52 &  19.46  $\pm$   0.22  &   17.98 $\pm$ 0.12 & 18.55 $\pm$  0.13  & --                 &  17.34 $\pm$  0.11 & & 1 \\ 
 2015-09-03 & 269.34 &  39.06 &   -- 		          &   18.07 $\pm$ 0.05 & 19.01 $\pm$  0.07  & 17.86  $\pm$ 0.04  &  17.26 $\pm$  0.06 & & 1 \\ 
 2015-09-03 & 269.34 &  39.06 &   --		          &   18.12 $\pm$ 0.07 & 18.90 $\pm$  0.06  & 17.77  $\pm$ 0.04  &  17.42 $\pm$  0.05 & & 1 \\ 
 2015-09-04 & 269.73 &  39.45 &  19.72  $\pm$   0.15  &   18.24 $\pm$ 0.05 & 18.94 $\pm$  0.06  & 17.72  $\pm$ 0.04  &  17.29 $\pm$  0.04 & & 1 \\ 
 2015-09-04 & 269.74 &  39.46 &  19.54  $\pm$   0.11  &   18.24 $\pm$ 0.05 & 18.88 $\pm$  0.07  & 17.73  $\pm$ 0.05  &  17.35 $\pm$  0.04 & & 1 \\
 2015-09-05 & 271.59 &  41.31 &  19.95  $\pm$   0.05  &   18.37 $\pm$ 0.10 & 19.09 $\pm$  0.01  & 17.90  $\pm$ 0.01  &  17.46 $\pm$  0.06  & 17.71$\pm$0.09 & 3 \\ 
 2015-09-08 & 274.33 &  44.05 &  19.32  $\pm$   0.10  &   18.27 $\pm$ 0.04 & 19.04 $\pm$  0.06  & 17.86  $\pm$ 0.03  &  17.39 $\pm$  0.04 & & 1 \\ 
 2015-09-08 & 274.33 &  44.05 &  19.60  $\pm$   0.11  &   18.18 $\pm$ 0.05 & 19.14 $\pm$  0.06  & 17.78  $\pm$ 0.04  &  --  & 		    & 1 \\
 2015-09-08 & 274.34 &  44.06 &  19.97  $\pm$   0.16  &   18.27 $\pm$ 0.05 & 18.94 $\pm$  0.06  & 17.86  $\pm$ 0.05  &  --  & 		    & 1 \\
 2015-09-08 & 274.37 &  44.09 &  19.62  $\pm$   0.12  &   --               & 18.99 $\pm$  0.06  & 17.70  $\pm$ 0.08  &  --  & 		    & 1 \\
 2015-09-08 & 274.38 &  44.10 &  19.74  $\pm$   0.14  &   --               & 18.96 $\pm$  0.07  & 17.80  $\pm$ 0.04  &  --  & 		    & 1 \\
 2015-09-08 & 274.47 &  44.19 &  20.00  $\pm$   0.48  &   18.00 $\pm$ 0.02 & 18.73 $\pm$  0.01  & 18.03  $\pm$ 0.02  &  17.60 $\pm$  0.03  & 17.76$\pm$0.06& 2 \\
 2015-09-09 & 274.52 &  44.24 &  19.48  $\pm$   0.13  &   --               & --                 & --                 &  17.37 $\pm$  0.04 & & 1 \\  
 2015-09-09 & 274.53 &  44.25 &  19.42  $\pm$   0.10  &   --               & --                 & --                 &  --  & & 1 \\
 2015-09-17 & 282.78 &  52.50 &  19.71  $\pm$   0.06  &   18.43 $\pm$ 0.03 & 19.07 $\pm$  0.03  & 18.01  $\pm$ 0.03  &  17.57 $\pm$  0.04 & & 1 \\ 
 2015-09-17 & 282.79 &  52.51 &  19.77  $\pm$   0.06  &   18.37 $\pm$ 0.03 & 19.18 $\pm$  0.03  & 18.01  $\pm$ 0.03  &  17.49 $\pm$  0.03 & & 1 \\
 2015-09-25 & 290.60 &  60.32 &  --                   &   --               & --                 & 18.25  $\pm$ 0.08  &  17.49 $\pm$  0.10 & & 1 \\ 
 2015-09-25 & 290.61 &  60.33 &  --                   &   --               & --                 & 18.10  $\pm$ 0.06  &  17.71 $\pm$  0.08 & & 1 \\ 
 2015-10-04 & 299.74 &  69.46 &  --                   &   18.50 $\pm$ 0.14 & 19.23 $\pm$  0.12  & --                 &  --  & & 1 \\
 2015-10-04 & 299.74 &  69.46 &  --                   &   18.65 $\pm$ 0.13 & --                 & --                 &  --  & & 1 \\
 2015-10-07 & 302.66 &  72.38 &  20.00  $\pm$   0.07  &   18.71 $\pm$ 0.03 & 19.35 $\pm$  0.03  & 18.38  $\pm$ 0.03  &  17.99 $\pm$  0.02 & & 1 \\ 
 2015-10-07 & 302.67 &  72.39 &  19.93  $\pm$   0.06  &   18.68 $\pm$ 0.03 & 19.29 $\pm$  0.03  & 18.37  $\pm$ 0.03  &  17.95 $\pm$  0.02 & & 1 \\ 
 2015-10-16 & 311.70 &  81.42 &  20.08  $\pm$   0.08  &   18.87 $\pm$ 0.04 & 19.41 $\pm$  0.04  & 18.50  $\pm$ 0.03  &  18.22 $\pm$  0.03 & & 1 \\ 
 2015-10-16 & 311.70 &  81.42 &  19.99  $\pm$   0.09  &   18.77 $\pm$ 0.04 & 19.39 $\pm$  0.04  & 18.53  $\pm$ 0.03  &  18.24 $\pm$  0.03 & & 1 \\ 
 2015-10-26 & 321.51 &  91.23 &  20.42  $\pm$   0.31  &   19.03 $\pm$ 0.15 & --  		& 18.64  $\pm$ 0.11  &  18.65 $\pm$  0.14 & & 1 \\ 
 2015-10-26 & 321.52 &  91.24 &  20.17  $\pm$   0.22  &   19.33 $\pm$ 0.15 & --                 & 18.81  $\pm$ 0.13  &  18.37 $\pm$  0.16 & & 1 \\ 
 2015-11-03 & 329.55 &  99.27 &  20.64  $\pm$   0.13  &   19.30 $\pm$ 0.06 & 19.75 $\pm$  0.05  & 18.94  $\pm$ 0.05  &  18.49 $\pm$  0.04 & & 1 \\ 
 2015-11-03 & 329.56 &  99.28 &  20.58  $\pm$   0.14  &   19.31 $\pm$ 0.06 & 19.89 $\pm$  0.05  & 18.93  $\pm$ 0.06  &  18.51 $\pm$  0.03 & & 1 \\
 2015-11-04 & 331.43 & 101.16 &  20.58  $\pm$   0.03  &   19.20 $\pm$ 0.02 & 20.08 $\pm$  0.03  & 18.89  $\pm$ 0.01  &  18.28 $\pm$  0.07   &18.12$\pm$0.11 & 3 \\ 
 2015-11-08 & 335.34 & 105.06 &  20.02  $\pm$   0.14  &   19.34 $\pm$ 0.01 & 19.69 $\pm$  0.02  & 19.33  $\pm$ 0.01  &  18.67 $\pm$  0.01   &18.20$\pm$0.06 & 2 \\ 
 2015-11-12 & 338.52 & 108.24 &  --                   &   --               & 20.06 $\pm$  0.09  & 18.95  $\pm$ 0.05  &  18.61 $\pm$  0.05 & & 1 \\ 
 2015-11-12 & 338.53 & 108.25 &  --                   &   --               & 19.85 $\pm$  0.06  & 18.94  $\pm$ 0.05  &  18.64 $\pm$  0.04 & & 1 \\ 
 2015-11-14 & 340.53 & 110.25 &  20.79  $\pm$   0.15  &   19.48 $\pm$ 0.07 & 19.88 $\pm$  0.05  & 19.05  $\pm$ 0.03  &  18.65 $\pm$  0.05 & & 1 \\ 
 2015-11-14 & 340.53 & 110.25 &  20.54  $\pm$   0.11  &   19.38 $\pm$ 0.08 & 19.87 $\pm$  0.05  & 19.05  $\pm$ 0.06  &  18.69 $\pm$  0.04 & & 1 \\ 
 2015-11-22 & 348.60 & 118.32 &  --                   &   19.44 $\pm$ 0.15 & 20.24 $\pm$  0.19  & 19.03  $\pm$ 0.12  &  18.68 $\pm$  0.13 & & 1 \\ 
 2015-11-22 & 348.61 & 118.33 &  --                   &   --               & 19.54 $\pm$  0.12  & 19.25  $\pm$ 0.12  &  18.80 $\pm$  0.13 & & 1 \\ 
 2015-11-30 & 356.54 & 126.26 &  20.73  $\pm$   0.15  &   19.55 $\pm$ 0.06 & 20.13 $\pm$  0.06  & 19.27  $\pm$ 0.06  &  18.88 $\pm$  0.07 & & 1 \\ 
 2015-11-30 & 356.55 & 126.27 &  20.74  $\pm$   0.17  &   19.59 $\pm$ 0.05 & 20.21 $\pm$  0.08  & 19.29  $\pm$ 0.05  &  18.92 $\pm$  0.04 & & 1 \\
 2015-12-02 & 335.34 & 128.97 &  20.56  $\pm$   0.19  &   19.36 $\pm$ 0.02 & 20.08 $\pm$  0.03  & 19.47  $\pm$ 0.02  &  18.88 $\pm$  0.03   & 18.72$\pm$0.06 & 2 \\ 
 2015-12-05 & 359.25 & 131.91 &  --                   &   19.76 $\pm$ 0.03 & 20.03 $\pm$  0.19  & 19.64  $\pm$ 0.03  &  19.00 $\pm$  0.02   & 18.89$\pm$0.07 & 2 \\  
 2015-12-09 & 362.53 & 135.25 &  20.75  $\pm$   0.24  &   19.74 $\pm$ 0.11 & 20.35 $\pm$  0.09  & 19.38  $\pm$ 0.07  &  19.06 $\pm$  0.07 & & 1 \\ 
 2015-12-09 & 365.53 & 135.25 &  21.21  $\pm$   0.26  &   19.80 $\pm$ 0.10 & 20.28 $\pm$  0.08  & 19.33  $\pm$ 0.06  &  --                & & 1 \\
 2015-12-10 & 366.53 & 136.25 &  20.73  $\pm$   0.18  &   19.78 $\pm$ 0.11 & 20.33 $\pm$  0.08  & 19.34  $\pm$ 0.09  &  19.02 $\pm$  0.08 & & 1 \\ 
 2015-12-10 & 366.54 & 136.26 &  21.01  $\pm$   0.22  &   19.88 $\pm$ 0.14 & 20.23 $\pm$  0.08  & --                 &  19.22 $\pm$  0.08 & & 1 \\
 2015-12-19 & 376.23 & 145.95 &  --                   &   20.01 $\pm$ 0.02 & 20.34 $\pm$  0.03  & 19.77  $\pm$ 0.02  &  19.22 $\pm$  0.03 & 19.02$\pm$0.07 & 2 \\ 
 2016-08-08 & 609.57 & 379.29 &  --                   &   --               & --                 & 20.88  $\pm$ 0.04  &   --               & --             & 2 \\

\hline    
\end{tabular}
\newline
$^\dagger$ JD 2,457,000+ ,
$^\ddagger$ Phase has been calculated with respect to V$_{max}$ =2457230.28   
$^\star$ 1: 1m LCO, 2: 1.82m Copernico Telescope, 3: 2.56m Nordic Optical Telescope (NOT)                                                                              
\label{tab:photometric_observational_log}                                                        
\end{table*} 

Wavelength and flux calibrations of all spectra acquired with the FLOYDS spectrograph were done using the \texttt{floydsspec}\footnote{https://www.authorea.com/users/598/articles/6566} pipeline. The spectra obtained from the 1.82m Copernico telescope and the 2.56m NOT were reduced using standard IRAF packages. The spectra were then wavelength and flux calibrated following standard steps.  The log of spectroscopic observations is given in Table \ref{tab:spectroscopic_observations}.  

\section{Explosion epoch, distance and Extinction}
\label{Explosion epoch, distance and Extinction}

  To ascertain the explosion epoch we have performed template fitting to the light curve of SN 2015dj using well constrained light curves of other type Ib SNe and found a best match with SN 1999ex. Template fitting involves applying time shift and magnitude scaling to the template light curves to best-fit the observed light curves. We have used  $\chi^{2}$ minimization to get the time of explosion of SN 2015dj by using {\it V} band light curve (Figure \ref{fig:light_curve}). The estimated value of explosion epoch from template fit is JD = 2457209.58$\pm$2 (consistent with the calculation given in section \ref{Temporal evolution of SN 2015dj}).
 
\citet{1981ApJS...45..541P} and \cite{1988cng..book.....T} quoted two measurements, 30.0 and 31.2 Mpc, respectively, for the host galaxy distance. The two estimates are a result from the different values of $H_0$ used. On the other hand, the luminosity distance is $D_L = 36.69\pm0.05$ Mpc for a redshift z = 0.008950$\pm$0.000013 (\cite{2004AJ....128...16K}) of NGC 7371, assuming $H_0$ = 73$\pm$5 km s$^{-1}$ Mpc$^{-1}$, $\Omega_m$ = 0.27,  $\Omega_v$ = 0.73). We have adopted the luminosity distance throughout the paper.

The Galactic extinction along the line of sight of SN 2015dj is $E(B-V) = 0.0498 \pm 0.0016$ mag \citep{2011ApJ...737..103S}.  The presence of NaI D lines in the SN spectra at the redshift of the host galaxy is indicative of some additional host extinction. The equivalent width (EW) of the NaI D feature is empirically related to the $E(B-V)$ value \citep{2012MNRAS.426.1465P}. The best signal-to-noise ratio (SNR) spectra of SN 2015dj taken on 2015 August 13 and 2015 September 05 show  a clear NaI D with EWs of 0.95 $\pm$ 0.06 \AA~ and 0.86$\pm$ 0.35 \AA, respectively. Using the weighted average of the two EWs (0.94 $\pm$ 0.06 \AA), and following \cite{1997A&A...318..269M} and \cite{2012MNRAS.426.1465P}, we obtain $E(B-V)_{host}$ = 0.24 $\pm$ 0.02 mag and 0.18 $\pm$ 0.01 mag, respectively. We adopt $E(B-V)_{host}$ = 0.19 $\pm$ 0.01 mag which is the weighted mean of the above two values. Combining the Galactic and host reddening contributions, the inferred value of $E(B-V)_{total}$ = 0.24 $\pm$ 0.01 mag, which is used throughout the paper.

\section{Photometric evolution of SN 2015dj}
\label{Temporal evolution of SN 2015dj}

\begin{figure}
	\begin{center}
		\includegraphics[width=\columnwidth]{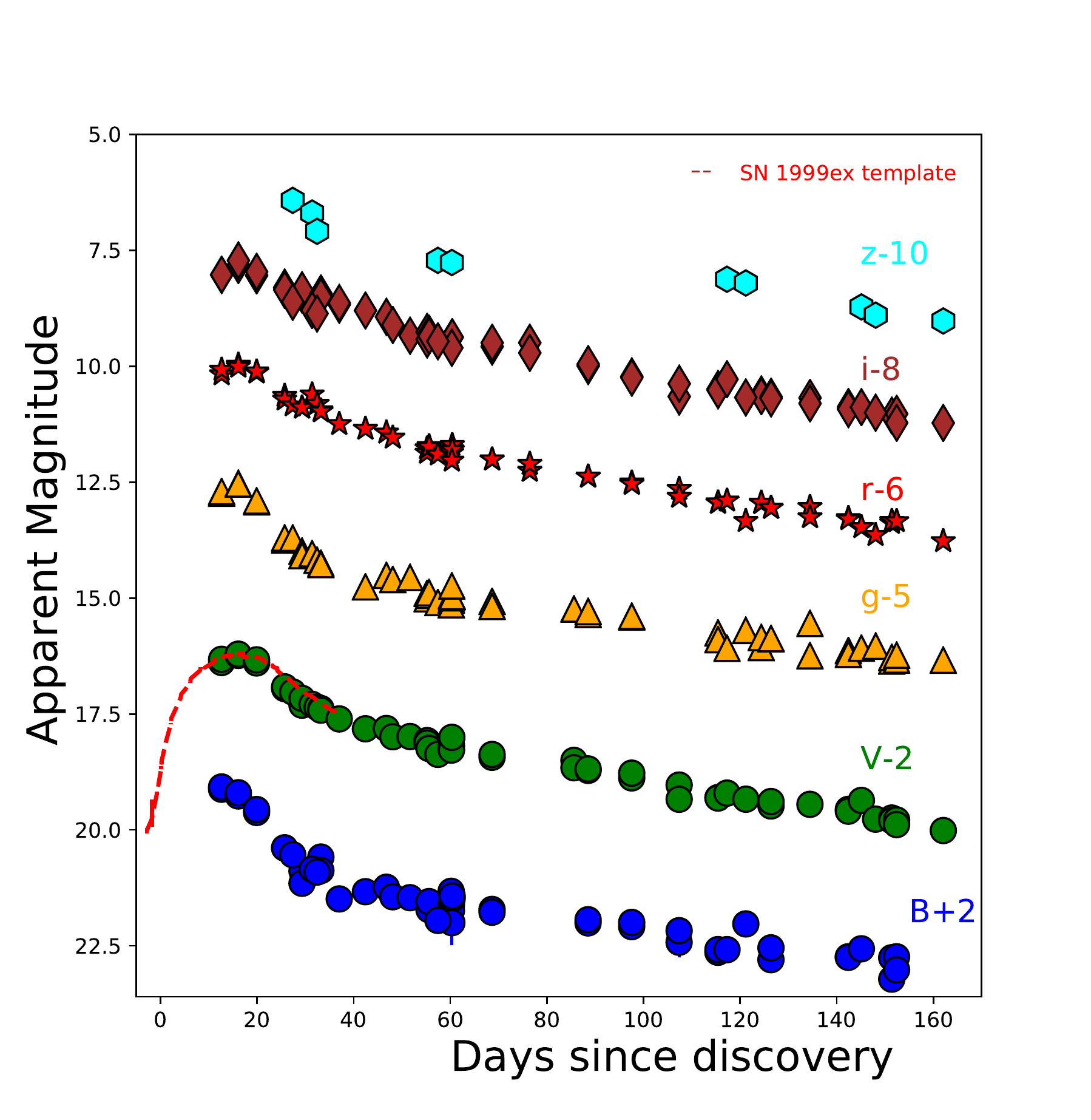}
	\end{center}
	\caption{Broadband {\it BVgriz} light curves of SN 2015dj. The {\it V} band light curve of the best fitting template SN 1999ex is over-plotted on the {\it V} band light curve of SN 2015dj with red dashed lines. An arbitrary offset has been applied in all bands for clarity.}
	\label{fig:light_curve}
\end{figure}

Figure \ref{fig:light_curve} shows the {\it BVgriz} light curves of SN 2015dj. The photometric observations started $\sim$ 13 days after discovery and lasted $\sim$ 170 days. The SN was also detected at $\sim$ 395 days after discovery in the $r$ band. The peak is well sampled in the {\it Vgri} bands. The peak magnitudes listed in Table \ref{tab:decay rate} are estimated using a cubic spline fit to the {\it Vgri} band light curves.  We use interpolation around the peak to estimate the associated errors in these measurements. The time of $V_{max}$ is used as a reference epoch for all the light curves.  The decay rates of SN 2015dj at different time intervals are reported in Table \ref{tab:decay rate}. 

\begin{figure}
	\begin{center}
		\includegraphics[width=\columnwidth]{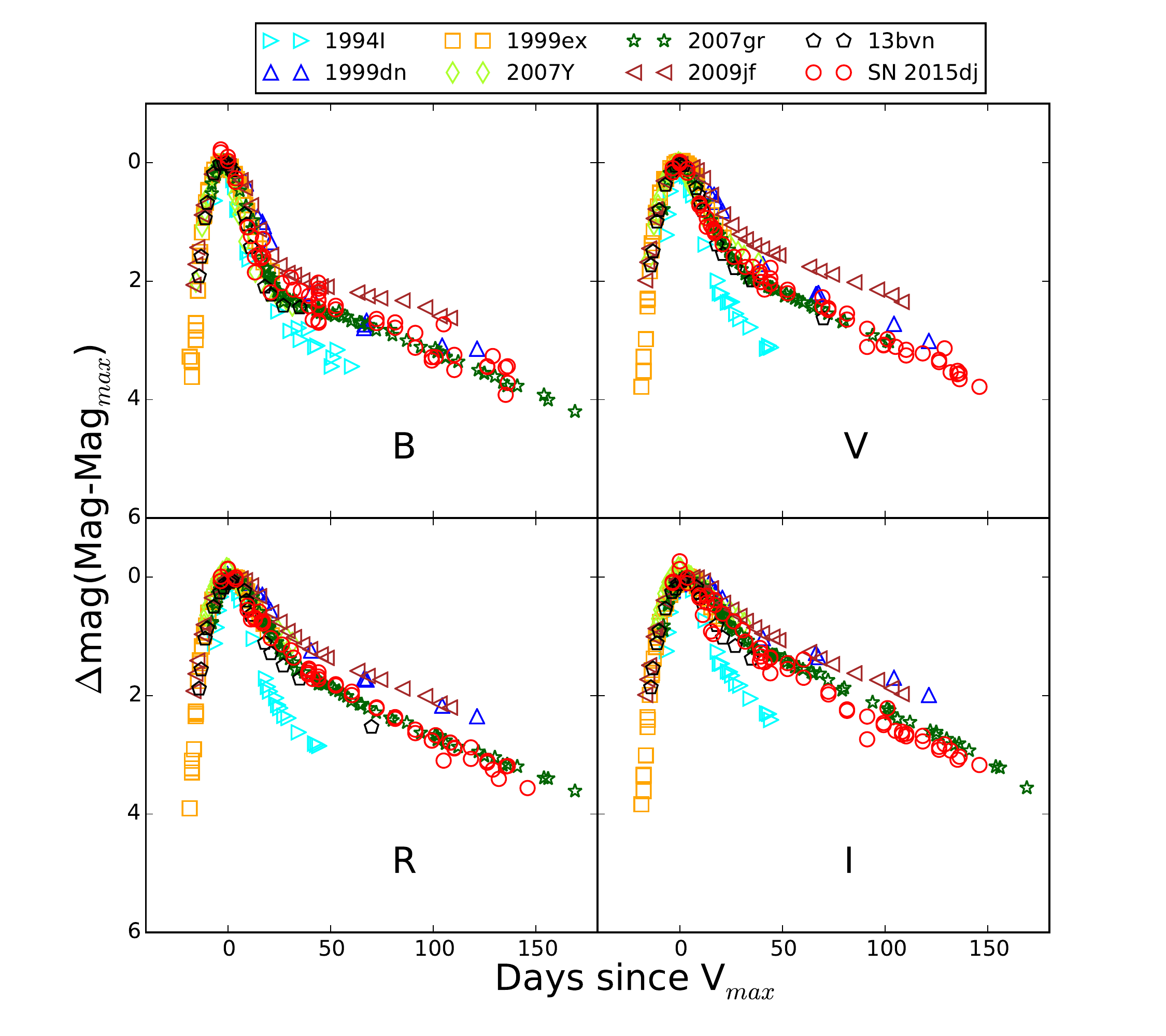}
	\end{center}
	\caption{Comparison of light curves of SN 2015dj with other type Ib/Ic SNe. The X-axis is the time from the {\it V}-band maximum, the Y-axis is the apparent SN magnitudes normalized to their magnitude at maximum.}
	\label{fig:comparison_light_curve.pdf}
\end{figure}

\begin{table*}
\caption{Log of spectroscopic observations}
\centering
\smallskip
\begin{tabular}{c c c c c c}
\hline \hline
Date          & Phase$^\dagger$          & Grism      & Spectral Range  & Resolution      & Telescope       \\
              &(Days)                    &            & (\AA)           &                 &                 \\
\hline
2015-07-10      & -16.10    & Gr04          & 3300-11000   & 311	      & AFOSC,EKAR  \\
2015-07-23      &  -3.70    & --       	  & 3200-10000   & 400-700     & FLOYDS, FTN  \\
2015-07-30      &  3.50     & --       	  & 3200-10000   & 400-700     & FLOYDS, FTN  \\
2015-08-07      & 10.70     & Gr04/VPH6 	  & 3300-11000   & 311/500 & AFOSC, EKAR \\
2015-08-08      & 12.10     & --       	  & 3200-10000   & 400-700     & FLOYDS, FTN   \\
2015-08-11      & 14.70     & Gr04,VPH6     & 3300-11000   & 311,500   & AFOSC, EKAR   \\
2015-08-13      & 16.80     & Gr04,VPH6     & 3300-11000   & 311,500    & AFOSC, EKAR    \\
2015-08-16      & 20.40     & --       	  & 3200-10000   & 400-700     & FLOYDS, FTN     \\
2015-08-28      & 32.20     & --            & 3200-10000   & 400-700     & FLOYDS, FTN   \\
2015-09-05      & 40.10     & --       	  & 3200-10000   & 400-700     & FLOYDS, FTN      \\
2015-09-05      & 41.34     & Gr04          & 3200-9600    &  360        & ALFOSC,NOT    \\
2015-09-08      & 44.14     & Gr04/VPH6     & 3300-11000   & 311/500    & AFOSC, EKAR     \\
2015-09-16      & 51.20     & --       	  & 3200-10000   & 400-700     & FLOYDS, FTN      \\
2015-10-02      & 67.10     & --       	  & 3200-10000   & 400-700     & FLOYDS, FTN      \\
2015-10-19      & 84.10     & --      	  & 3200-10000   & 400-700     & FLOYDS, FTN      \\
2015-11-04      & 101.10    & Gr04          & 3200-9600    & 360         & ALFOSC,NOT    \\
2015-12-02      & 128.40    & Gr04          & 3200-9600    &  360        & ALFOSC, NOT    \\
2015-12-03      & 129.50    & VPH6          & 3300-11000   & 500        & AFOSC, EKAR     \\
2015-12-09      & 134.90    & --       	  & 3200-10000   & 400-700     & FLOYDS, FTN     \\
2015-12-29      & 154.90    & --       	  & 3200-10000   & 400-700     & FLOYDS, FTN     \\
\hline                                   
\end{tabular}
\newline
$^\dagger$ Phase is from V$_{max}$= 2457230.28 
\label{tab:spectroscopic_observations}      
\end{table*}

We compare the properties of SN 2015dj with those of other well-studied type Ib/Ic SNe available in the literature (Table \ref{tab:Properties of the comparison sample}): SNe 1994I ($E(B-V) = 0.452$ mag; $D_L = 6.3$ Mpc;  \cite{1996AJ....111..327R}), 1999dn ($E(B-V) = 0.100$ mag; $D_L = 38.6$ Mpc; \cite{2011MNRAS.411.2726B}), 1999ex ($E(B-V) = 0.300$ mag; $D_L = 47.3$ Mpc; \cite{2002AJ....124.2100S}), 2007Y ($E(B-V) = 0.112$ mag; $D_L = 19.1$ Mpc; \cite{2009ApJ...696..713S}), 2007gr ($E(B-V) = 0.092$ mag; $D_L = 7.1$ Mpc; \cite{2009A&A...508..371H}), 2009jf ($E(B-V) = 0.110$ mag; $D_L = 32.8$ Mpc; \cite{2011MNRAS.413.2583S}) and iPTF13bvn ($E(B-V) = 0.215$ mag; $D_L = 18.7$ Mpc; \cite{2014MNRAS.445.1932S}). These SNe were chosen to cover a wide range of luminosities ranging from the brightest (SN 2009jf) to the faintest (SN 2007Y). The sample includes proto typical type Ib/Ic SNe along with SNe having diverse properties than the normal type Ib/Ic SNe. For uniformity we adopt luminosity distance for all SNe in the comparison sample. Figure \ref{fig:comparison_light_curve.pdf} shows a comparison of SN 2015dj with other type Ib/c SNe. SN 2015dj has a slower decline than SN 1994I, whereas it declines faster than SNe 1999dn and 2009jf. An overall similarity is seen with the normal type Ic SN 2007gr in terms of the light-curve shape. 

\begin{table*}
\caption{ Parameters of SN 2015dj  }
\centering
\smallskip
\begin{tabular}{l  c c c c c}
\hline \hline
SN 2015dj                                      & V band & g band                & r band              & i band  \\
\hline
JD of maximum light (2457000+)                 & 230.3$\pm$0.5    & 230.3$\pm$0.5         & 230.3$\pm$0.5       &230.3$\pm$0.5              \\
Magnitude at maximum (mag)                   & 16.22$\pm$0.02   & 16.53$\pm$0.01        & 15.98$\pm$0.01      & 15.77$\pm$0.02    \\
Absolute magnitude at maximum (mag)          & -17.37$\pm$0.02  & -17.07$\pm$0.02       & -17.38$\pm$0.01     & -17.46$\pm$0.02    \\

Colour at maximum (mag)                 &                  &                       &                 &   	       \\
$(B-V)_{0}$                                  & 0.77 $\pm$ 0.08  &      		        &    	          &   	       \\
$(V-R)_{0}$                                  & 0.22 $\pm$ 0.05  &      		        &                 &   	       \\
$(V-I)_{0}$                                  & 0.43 $\pm$ 0.07  &      		        &                 &   	       \\
\hline 
Upper limits on SN magnitudes (mag)$\dagger$ & B band     & V band       & g band     & r band     & i band  \\
                                                & 22.39      &  21.87       & 22.59      & 21.69      & 21.34 \\
\hline
\hline
Decline rate $\ddagger$  & B band           & g band            & V band          & r band          & i band  \\
\hline
$\Delta$m$_{15}$( mag)                       &--                & 1.68$\pm$0.02     &  1.21$\pm$0.02  & 1.04$\pm$0.02   & 0.52$\pm$0.01\\                       
$\Delta m$(20 -- 60) mag (100 days)$^{-1}$ 	& 1.09$\pm$0.06     & 1.95$\pm$0.09     & 2.43$\pm$0.05    & 2.45$\pm$0.04   & 2.52$\pm$0.05 \\
$\Delta m$(60 -- 136) mag (100 days)$^{-1}$  	& 1.59$\pm$0.03     & 1.51$\pm$0.03     & 1.73$\pm$0.03    & 1.63$\pm$0.01   & 1.74$\pm$0.03   \\
\hline

\end{tabular}
\newline
$^\dagger$ Template images of LCO data\\
$^\ddagger$with respect to V$_{max}$= 2457230.28 
\label{tab:decay rate}      
\end{table*}

\begin{table*}
\caption{Properties of the comparison sample }
\centering
\smallskip
\begin{tabular}{c c c c c c c  }
\hline \hline
                       & SNe        & Host galaxy & Distance*   & Extinction  & M$_Ni$$^\ddagger$                   &  Reference$^\dagger$  \\
                       &            &             & (Mpc)       & E(B-V)   &M$_{\odot}$                                                              							 \\
\hline
                       & SN 1994I   & M51         & 6.3         & 0.452       &  0.06$\pm$0.01      &							1     \\
                       & SN 1999dn  & NGC 7714    & 38.6        & 0.100       &  0.04$\pm$0.01      &							2      \\
		               & SN 1999ex  & IC 5179     & 47.3        & 0.300       &  0.14$\pm$0.02 	    &							3      \\
                       & SN 2007Y   & NGC 1187    & 19.1        & 0.112       &  0.02               &							4      \\
		               & SN 2007gr  & NGC 1058    & 7.1         & 0.092       &  0.02               &							5     \\
                       & SN 2009jf  & NGC 7479    & 32.8        & 0.110       &  0.15$\pm$0.01      &							6     \\
                       & iPTF 13bvn & NGC 5806    & 18.7        & 0.215       &  0.03$\pm$0.01      &							7     \\
		       
\hline                                                                                   
\end{tabular}
\newline
$^\dagger$ REFERENCES.-- (1)\cite{1996AJ....111..327R}, NED; (2)\cite{2011MNRAS.411.2726B}, NED; (3)\cite{2002AJ....124.2100S}, NED; (4)\cite{2009ApJ...696..713S}, NED; (5)\cite{2009A&A...508..371H}, NED; (6)\cite{2011MNRAS.413.2583S}, NED; (7)\cite{2014MNRAS.445.1932S}, NED.
*Luminosity distance
$^\ddagger$These values are estimated using the same model adopted for SN 2015dj. 
\label{tab:Properties of the comparison sample}        
\end{table*}

Accounting for the distance and extinction adopted in Section \ref{Explosion epoch, distance and Extinction}, we compute the peak absolute magnitudes of SN 2015dj in the {\it Vgri} bands (Table \ref{tab:decay rate}). The $V$-band peak absolute magnitude of SN 2015dj is $M{_V} = -17.37\pm0.02$ mag, which is typical of a type Ib SN \citep{2018A&A...609A.136T}. 

\begin{figure}
	\begin{center}
		\includegraphics[width=\columnwidth]{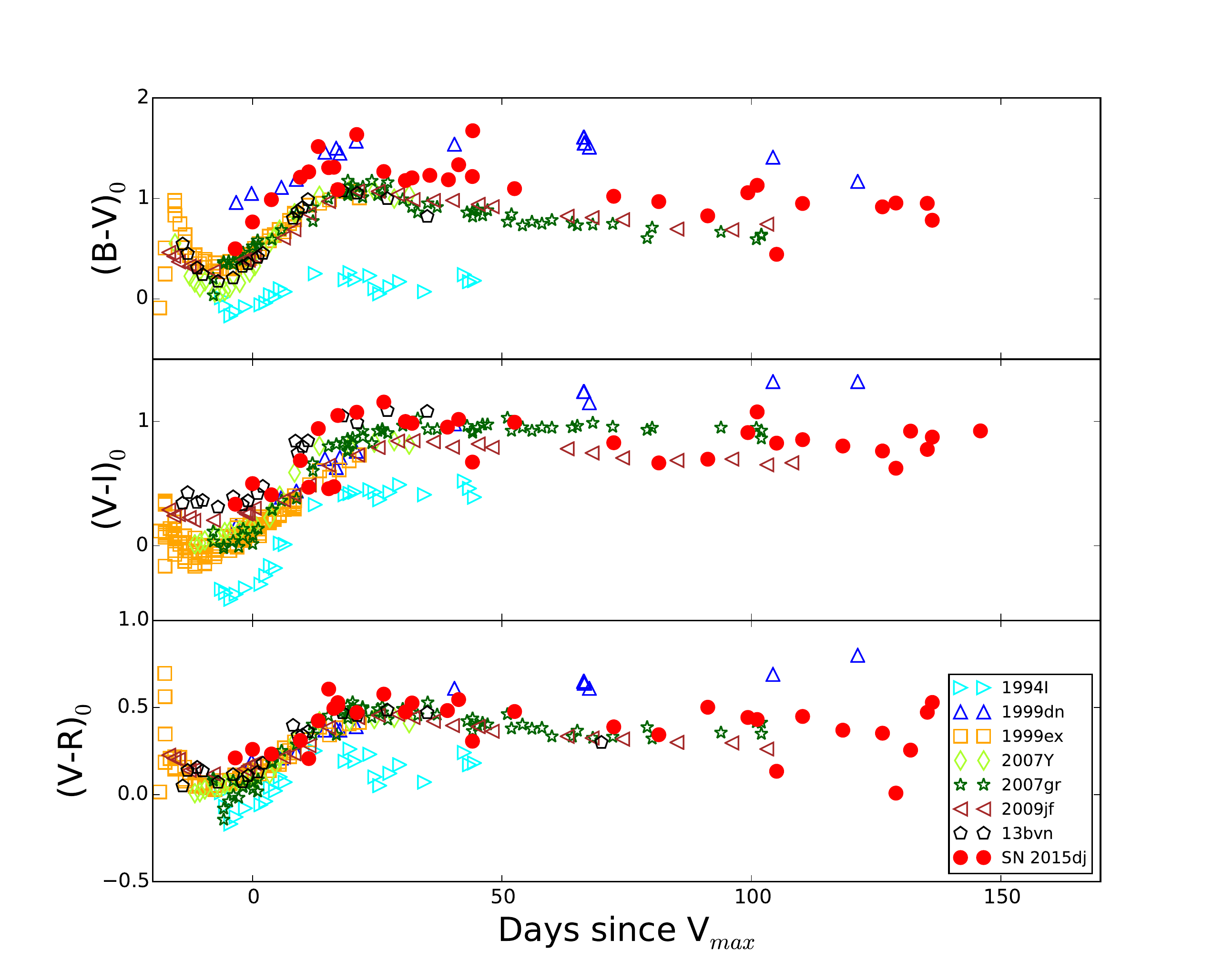}
	\end{center}
	\caption{Comparison of ($B-V$)$_{0}$, ($V-I$)$_{0}$ and ($V-R$)$_{0}$ colours of SN 2015dj with those of other type Ib/Ic SNe.}
	\label{fig:color_curve}
\end{figure}

A depiction of the colour evolution of SN 2015dj and its comparison with our SN Ib/Ic sample is shown in Figure \ref{fig:color_curve}. The ($B-V$)$_{0}$ colour of SN 2015dj reaches $\sim$ 0.5 mag at $\sim$ $-3$ days and 1.7 mag after $\sim$ 45 days.  A gradual decrease is noticed in the ($B-V$)$_{0}$ colour evolution. Initially the ($B-V$)$_{0}$ colour of SN 2015dj is bluer than SN 1999dn and redder than all other SNe. This trend also continues at intermediate and late phases. The $(V-I)_{0}$ colour rises to $\sim$ 1.1 mag between 10 and 40 days. A drop by 0.4 mag is observed from $\sim$ 50 days to 90 days. The ($V-I$)$_{0}$ colour is initially similar to iPTF13bvn and redder than other SNe. After $\sim$ 40 days, the colour evolution of SN 2015dj is similar to SN 2007gr. The ($V-R$)$_{0}$ colour increases by $\sim$ 0.3 mag at $\sim$ 40 days and remains constant up to $\sim$ 80 days. From 50 to 100 days, the ($V-R$)$_{0}$ colour of SN 2015dj is bluer than that of SN 1999dn, and redder than those of SNe 2007gr and 2009jf.

The $(V-R)_{0}$ colour is an independent tool to probe the reddening along the line of sight.  \citet{2011ApJ...741...97D} reported a mean value of $\langle$ ($V-R$)$_{V10}$ $\rangle$ = 0.26$\pm$0.06 mag. Here $\langle$ ($V-R$)$_{V10}$ $\rangle$ is the $(V-R)_{0}$ colour at 10 days since $V_{max}$. For SN 2015dj, we estimate $\langle$ ($V-R$)$_{V10}$~$\rangle$ to be 0.31$\pm$0.05 mag, in agreement with \citet{2011ApJ...741...97D}.

\begin{figure}
	\begin{center}
		\includegraphics[width=\columnwidth]{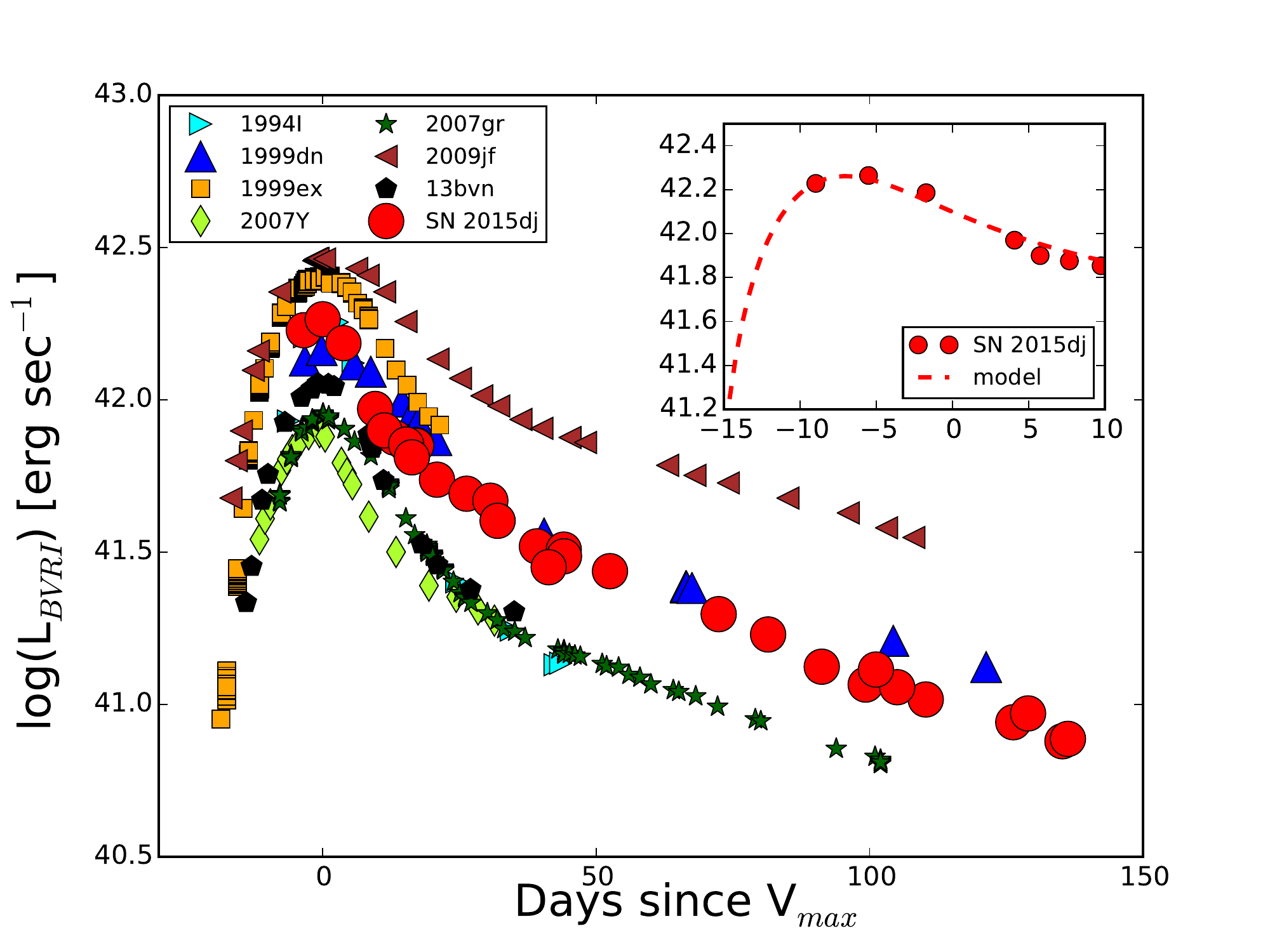}
	\end{center}
	\caption{{\it BVRI} pseudo bolometric light curve of SN 2015dj compared with those of other type Ib/c SNe. The inset shows the best fit model to the pseudo bolometric light curve of SN 2015dj.}
	\label{fig:final_bolometric_plot}
\end{figure}

We construct the pseudo-bolometric light curve of SN~2015dj using the extinction corrected magnitudes in the {\it BVRI} bands and the luminosity distance  (see Section \ref{Explosion epoch, distance and Extinction}). The fluxes, obtained by converting the magnitudes using the zero points given in \cite{1998A&A...333..231B}, are integrated using the trapezoidal rule between {\it B} and {\it I} bands. The integrated flux is converted to luminosity. The pseudo bolometric light curves of the comparison sample are constructed following the same method as SN 2015dj. In Figure \ref{fig:final_bolometric_plot}, the pseudo bolometric light curve of SN 2015dj is shown along with those of the comparison sample. SN 2015dj, with a peak pseudo bolometric luminosity of $\sim$ 1.8 $\times$ 10$^{42}$ erg s$^{-1}$, is fainter than SNe 1999ex and 2009jf, while it is brighter than all other comparison SNe.     

In order to estimate the parameters of explosion for SN 2015dj, such as the ejected mass ($M_{\rm ej}$), the 
$^{56}$Ni mass ($M_{\rm Ni}$) and the kinetic energy of the explosion ($E_{\rm k}$), we use the Arnett 
model \citep{1982ApJ...253..785A}, formulated by \cite{2008MNRAS.383.1485V}. It assumes spherical symmetry, homologous expansion, constant opacity, a small pre-explosion radius and no mixing of $^{56}$Ni in the ejecta. The lack of pre-maximum points in the light curve, and the simplified assumptions in the analytical model mean that the explosion parameters estimated here should be treated as an order of magnitude estimates.
The early photospheric phase light curve ($\lesssim 30$d past explosion) was used to perform the analysis.
The free parameters in the model are $\tau_{\rm m}$ (timescale of the light curve) and $M_{\rm Ni}$, where the former is 
defined as 

$$ \tau_{\rm m} = \left( \frac{\kappa_{\rm opt}}{\beta c} \right)^{1/2} \left( \frac{6 M_{\rm ej}^3}{5 E_{\rm k}} \right)^{1/4} .$$
\noindent
The kinetic energy is expressed as 
$$ E_{\rm k} = \frac{3}{5} \frac{M_{\rm ej} v_{\rm ph}^2}{2}$$

\noindent
where, $\beta \simeq 13.8$ is a constant of integration \citep{1982ApJ...253..785A} and $v_{\rm ph}$ is the photospheric  velocity, that can be determined observationally using the ejecta expansion velocity at maximum 
light \citep{2016MNRAS.457..328L}. We use a constant opacity $\kappa_{\rm opt} = 0.06$ cm$^2$ g$^{-1}$.

The best fit parameters obtained for a $v_{\rm ph} = 9,000$~km~s$^{-1}$ are $\tau_{\rm m} = 5.8 \pm 0.3$ days and 
$M_{\rm Ni} = 0.05 \pm 0.01$ M$_{\odot}$. Solving for the ejected mass and kinetic energy of the explosion, we obtain 
$M_{\rm ej} \sim 0.5$~M$_{\odot}$ and $E_{\rm k} \sim 0.3 \times 10^{51}$ erg. 
The fit favours a low rise time of $\sim 9$ days. The same model is adopted to fit the pseudo bolometric light curves of comparison SNe and conclude that SN 2015dj shares similarity with SNe 1994I and 1999dn in terms of $^{56}$Ni synthesized during explosion. The pseudo bolometric light curve along with the best fit model are shown in the inset of Figure \ref{fig:final_bolometric_plot}. 

We caution that the above estimates of the explosion parameters and rise time are based on a light curve that lacks coverage during the pre-maximum phase. In order to quantify the effect of missing pre-maximum (with respect to V-band) data on the inferred parameters, we fit our model to the objects in our comparison sample by progressively excluding pre-maximum data. The rise time is also allowed to vary in the fit, instead of adopting the values in the literature.  

The rise time for SN 2015dj inferred from the bolometric light curve is likely underestimated significantly. We find that the best-fit rise time, on average for our comparison objects, falls by a factor of 0.6 $\pm$ 0.1 when all of the pre-maximum data is excluded (relative to best-fit rise time for the full light curve). Based on this, we estimate a `true' rise time for SN 2015dj of  $15\pm 5$ days. 
If the radioactive material is concentrated in the inner regions of the ejecta, the SN could show a `dark phase' of several days after the explosion \citep{2013ApJ...769...67P}, adding further uncertainty to the rise time inferred from the bolometric light curve.
From the bolometric light curve we have a lower limit on the rise time of ~9 days, whereas a deep Pan-STARRS1 w-band non-detection (w $\geq$ 21.8) places an upper limit of ~30 days.

The $M_{\rm Ni}$ and $\tau_{\rm m}$ (and thus $M_{\rm ej}$ and $E_{\rm k}$) values estimated above are also considered limits, with `true' values revised to $M_{\rm Ni} = 0.06 \pm 0.01$ \msol, $\tau_{\rm m} = 9.5^{+3.6}_{-2.1}$ days, $M_{\rm ej} = 1.4^{+1.3}_{-0.5}$ \msol\, and $E_{\rm k} = 0.7^{+0.6}_{-0.3} \times 10^{51}$ erg.

\section{Spectral Evolution}
\label{spectral_evolution}

Our spectroscopic campaign started on the day of discovery and lasted $\sim$ 160 days. In the following sections, we will describe the spectral evolution of SN 2015dj close to peak (Section \ref{Pre-maximum spectral phase}), after maximum (Section \ref{Post-maximum spectral phase}) and in the nebular phase (Section \ref{nebular spectral phase}). The spectra presented in the figures have been all redshift and reddening corrected. 

\subsection{Pre-maximum spectra}
\label{Pre-maximum spectral phase}

\begin{figure*}
	\begin{center}
		\includegraphics[width=\textwidth]{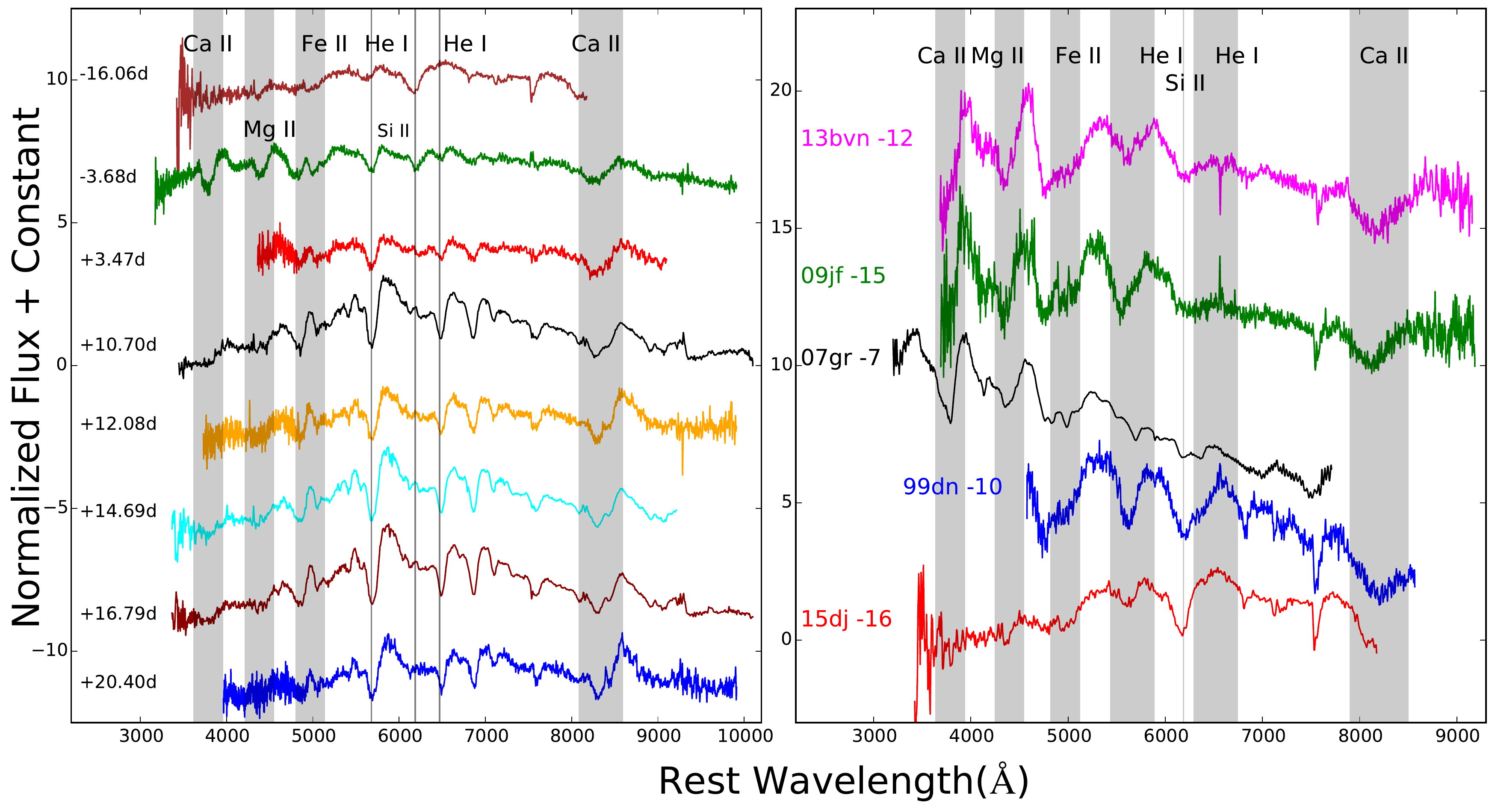}
	\end{center}
	\caption{{\bf Left panel:} Spectral evolution of SN 2015dj from $-16$ days to 20 days from the maximum light. The different lines are marked with shaded bars. {\bf Right panel:} Comparison of pre maximum spectral features of SN 2015dj with those of other type Ib/Ic SNe.}
	\label{fig:SN 2015dj_spectral_sequence_subplot_1}
\end{figure*}

Figure \ref{fig:SN 2015dj_spectral_sequence_subplot_1} (left panel) presents the spectral evolution of SN 2015dj until $\sim$ 20 days since V$_\textrm{max}$. The pre-maximum spectra from $-16.1$ and $-3.7$ days are also shown. The most identifying feature is He I at 5876 \AA, which has a velocity of $\sim$ 13,500 km s$^{-1}$ at phase $-16.1$ days. The spectrum at $-3.7$ day shows strong P-Cygni profiles of Ca {\sc ii}, Mg {\sc ii}, Fe {\sc ii} and Si {\sc ii} lines which are marked with shaded bars in the figure. He I features at 5876 \AA, 6678 \AA~ and 7065 \AA~ are clearly detected. He I 5876 \AA~ has a velocity of $\sim$ 10,500 km s$^{-1}$, whereas the Ca {\sc ii} H$\&$K feature found in the blue region has a velocity of $\sim$ 12,000 km s$^{-1}$. The Fe {\sc ii} multiplet near 5000 \AA, and Ca {\sc ii} NIR features are also well detected in the spectrum. The absorption trough at $\sim$ 6200 \AA~ in the early spectra is due to the photospheric Si {\sc ii} with velocities $\sim$ 9000 km s$^{-1}$ and 7300 km s$^{-1}$ at phases $\sim$ $-16.1$ and $-3.7$ days, respectively. The blackbody temperature associated with first five spectra of SN 2015dj varies between 5200 and 4300 K.

Figure \ref{fig:SN 2015dj_spectral_sequence_subplot_1} (right panel) shows a pre-maximum spectral comparison of SN 2015dj  with other type Ib/Ic SNe, such as SNe 1999dn \citep{2000ApJ...540..452D,2011MNRAS.411.2726B}, 2007gr \citep{2008ApJ...673L.155V,2014AJ....147...99M}, 2009jf \citep{2011MNRAS.413.2583S,2014AJ....147...99M} and iPTF13bvn \citep{2014MNRAS.445.1932S}. At this phase, all spectral lines have P-Cygni profiles. SN 2015dj has narrower P-Cygni profiles, which is indicative of lower expansion velocity of the ejecta. Most of the identified lines are marked with shaded bars. The Fe {\sc ii} multiplet near 5000~\AA{} is detected in spectra of all SNe although that observed in SN 2015dj shares more similarity with SNe 1999dn and iPTF13bvn. The absorption profile of He I 5876~\AA{} is similar to that of iPTF13bvn, whereas the absorption of the Si~{\sc ii} line is similar to that of SN 1999dn. We notice an overall similarity between SN 2015dj and SN 1999dn.  

\subsection{Post-maximum spectra}
\label{Post-maximum spectral phase}

\begin{figure*}
	\begin{center}
		\includegraphics[width=\textwidth]{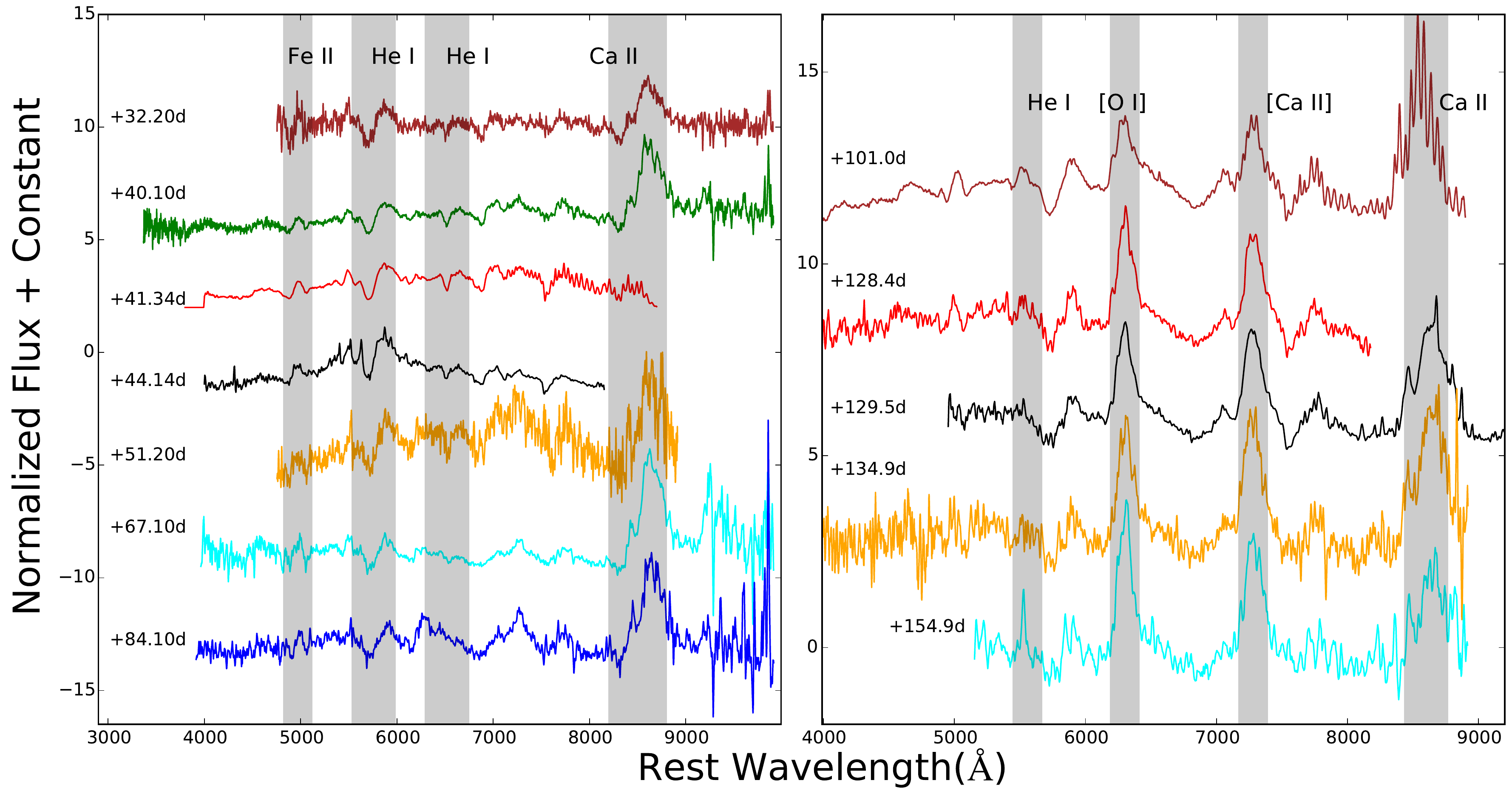}
	\end{center}
	\caption{{\bf Left panel:} Spectral evolution of SN 2015dj from 32 to 84 days post maximum. {\bf Right panel:} Spectral evolution of SN 2015dj from 101 to 154 days post maximum. Shaded bars represent different spectral features.}
	\label{fig:SN 2015dj_spectral_sequence_subplot_2}
\end{figure*}

Figure \ref{fig:SN 2015dj_spectral_sequence_subplot_2} (left panel) shows the spectral sequence of SN 2015dj from $\sim$ 32 days to $\sim$ 84 days after V$_{max}$. He~{\sc i}~6678~\AA~becomes fainter than He~{\sc i}~5876~\AA, whereas He~I~7065~\AA{} feature is very weak. The Ca~{\sc ii} NIR feature becomes dominated by the emission component at late phases. 
A comparison of the spectral features of SN 2015dj with other type Ib/c SNe at 10 day from V$_{max}$ is shown in left panel of Figure \ref{fig:comparison_subplot} (SNe 1999dn, \citealt{2000ApJ...540..452D,2011MNRAS.411.2726B}; 2007gr,  \citealt{2008ApJ...673L.155V,2014AJ....147...99M, 2019MNRAS.482.1545S};  2009jf, \citealt{2011MNRAS.413.2583S,2014AJ....147...99M, 2019MNRAS.482.1545S}; iPTF13bvn,  \citealt{2014MNRAS.445.1932S}). Ca~{\sc ii}~H \& K feature is well developed in all SNe except iPTF13bvn which lacks spectral coverage in that region. The Fe~{\sc ii} multiplet around 5000~\AA{} is present in all SNe, and this absorption feature of SN 2015dj is similar to that in SNe 2009jf and iPTF13bvn. He~I~5876~\AA, 6678~\AA, and 7065~\AA{} are well-developed in SN 2015dj at this epoch. The Fe~{\sc ii} multiplet near 5400~\AA{} is well developed in SN 2015dj as well as in SNe 2007gr and 2009jf. A weak absorption dip is seen in SNe 1999dn and iPTF13bvn. Ca~{\sc ii} NIR feature is well developed in SN 2015dj and almost similar to SNe 2009jf and 13bvn.

\subsection{Nebular phase spectra}
\label{nebular spectral phase}

Figure \ref{fig:SN 2015dj_spectral_sequence_subplot_2} (right panel) shows the spectral evolution from $\sim$ 101 to 154 days after V$_{max}$. At these late epochs, the SN enters the nebular phase and the spectral lines are now mostly in emission. [O~{\sc i}] 6300, 6364 \AA{} and [Ca~{\sc ii}] 7291, 7324 \AA{} are now clearly visible. If there would have been any presence of H$\alpha$ feature in the early time spectra of SN 2015dj, then [O~{\sc i}] profile should be associated with a redward broad shoulder \citep{2010MNRAS.409.1441M}. In the late time spectra of SN 2015dj we find no evidence of such redward broad component in [O~{\sc i}] profile. This is another evidence for SN 2015dj to be a type Ib SN. Although \cite{2016ApJ...820...75P} have discussed about equal possibilities of Si~{\sc ii} and H$\alpha$ line near 6200 \AA, late time spectral features rule out the possibility of H$\alpha$ signature in SN 2015dj. 

\begin{figure*}[!htbp]
		\includegraphics[width=\textwidth]{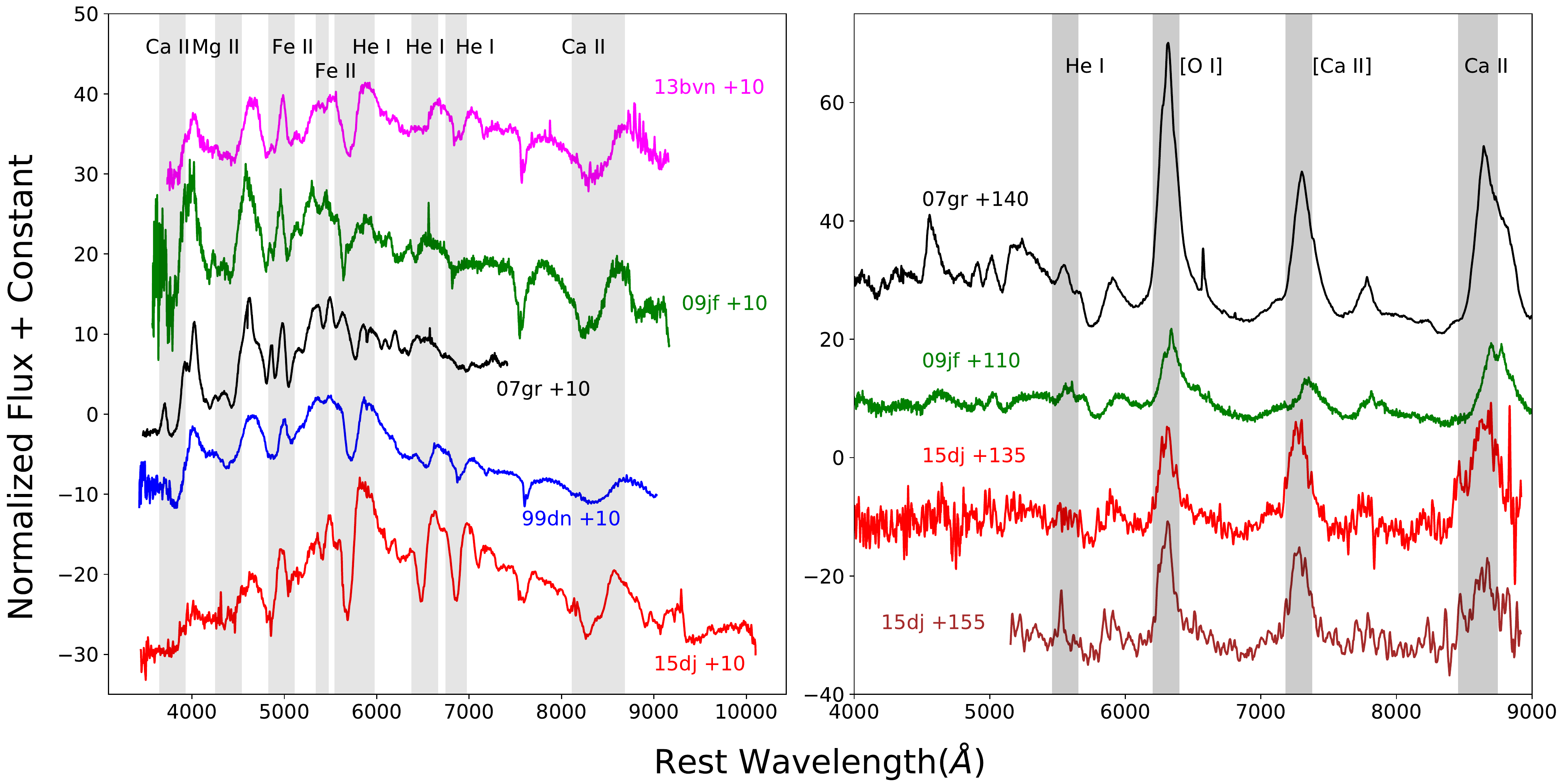}
	\caption{{\bf Left panel:} Comparison of post maximum ($\sim$10 day) and {\bf Right panel:} nebular phase spectra of SN 2015dj with those of other well studied type Ib/Ic SNe.}
	\label{fig:comparison_subplot}
\end{figure*}

In Figure \ref{fig:comparison_subplot} (right panel), we compare the nebular spectral features (from $\sim$ 135 to 155 days) of SN 2015dj with SNe 2007gr and 2009jf. Mg~{\sc i}] feature at 4571 \AA~ is present in SN 2007gr but is not significant in SNe 2009jf and 2015dj. The prominent [O~{\sc i}] doublet at 6300 \AA~ and 6364 \AA~ seen in the nebular phase spectrum is stronger in SN 2007gr as compared to SNe 2009jf and 2015dj, however, SN 2015dj bears a close resemblance in overall shape with SN 2009jf. At the red wavelengths, the Ca~{\sc ii} NIR feature shows a single sharp emission peak in SN 2007gr, double peak kind of structure in SN 2009jf and multipeaked asymmetric structure in SN 2015dj.

\subsection{Ejecta velocity evolution}
\label{Velocity Evolution}

\begin{figure}
	\begin{center}
		\includegraphics[width=\columnwidth]{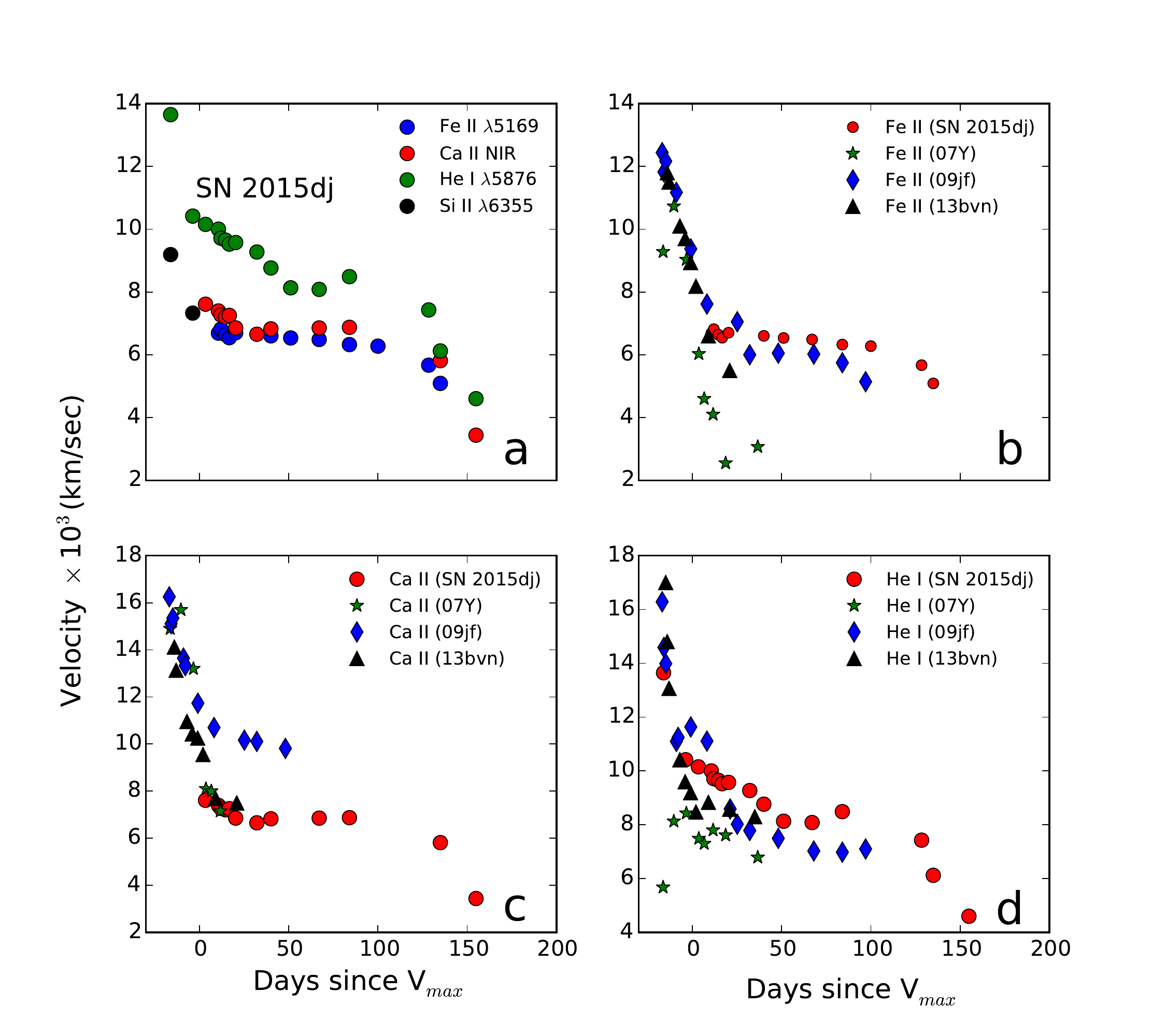}
	\end{center}
	\caption{Expansion velocity at different epochs for a few lines in SN 2015dj and other type Ib/Ic SNe.}
	\label{fig:velocity_plot_SN 2015dj}
\end{figure}

We measure the line velocities of Fe~{\sc ii}~5169~\AA, Si~{\sc ii}~6355~\AA, He~{\sc i}~5876~\AA{} and NIR Ca~{\sc ii} in SN 2015dj by fitting a Gaussian to the minimum of their P~Cygni absorption profiles (Figure \ref{fig:velocity_plot_SN 2015dj}, panel a). These lines originate in different temperature and density conditions with the corresponding elements being stratified within the SN ejecta. In general, the Ca~{\sc ii} lines show high velocities as they form in the outer ejecta, whereas the Fe line forming region is situated in the inner part of the ejecta. The Fe~{\sc ii} lines provide a good representation of the photospheric velocity. The Fe~{\sc ii}~5169~\AA{} velocity ranges between 7000~km~s$^{-1}$ and 4960~km~s$^{-1}$ from 10 to 134 days. The expansion velocities of Si~{\sc ii}~6355~\AA{} are $\sim$ 9000~km~s$^{-1}$ and 7300~km~s$^{-1}$ at $-16$ and $-3$ days from V$_{max}$, respectively. The velocity of He~{\sc i} line at 5876~\AA{} declines from $\sim$ 13,500~km~s$^{-1}$ at phase $\sim-16$ day to $\sim$ 4500~km~s$^{-1}$ at 155 day. 

Figure \ref{fig:velocity_plot_SN 2015dj} (panel b) compares the velocity evolution of the Fe~{\sc ii}~5169~\AA{} line in SN 2015dj and other type Ib/Ic SNe. The Fe~{\sc ii} line velocities for SN 2015dj are lower than SN 2009jf, but higher than those of SNe 2007Y and iPTF13bvn from $\sim$ 10 to 36 days. The velocities in SN 2015dj are nearly 500~km~s$^{-1}$ higher than those of SN 2009jf at $\sim$ 40-85 days. The velocity of the NIR Ca~{\sc ii}  feature in SN 2015dj is lowest, as compared to other SNe (Figure \ref{fig:velocity_plot_SN 2015dj}, panel c). The formation of Ca lines at unusually low velocity can be associated with large mixing in the ejecta \citep{2014MNRAS.445.1932S}. It is nearly 3500~km~s$^{-1}$ lower than that measured in SN 2009jf at phase $\sim$ 30 day, and is similar to those of SNe 2007Y and iPTF13bvn. The Ca~{\sc ii} NIR triplet velocity follows a flat evolution until $\sim$ 85 days and declines at later phases. At early epochs ($\sim$ $-$5 to +9 days) the He~{\sc i}~5876~\AA{} velocities in SN~2015dj are higher than iPTF13bvn and SN 2007Y but lower than SN 2009jf (Figure \ref{fig:velocity_plot_SN 2015dj}, panel d). From $\sim$ 34 days past maximum, the He~{\sc I} velocities are higher in SN 2015dj than other comparison SNe. Overall, we find that the velocity evolution of a few lines in SN 2015dj is similar to those of a normal type Ib SNe.

\subsection{Nebular phase [OI] and [Ca II] emission lines} 
\label{Nebular phase calcium emission}     

\begin{figure}
	\begin{center}
		\includegraphics[width=\columnwidth]{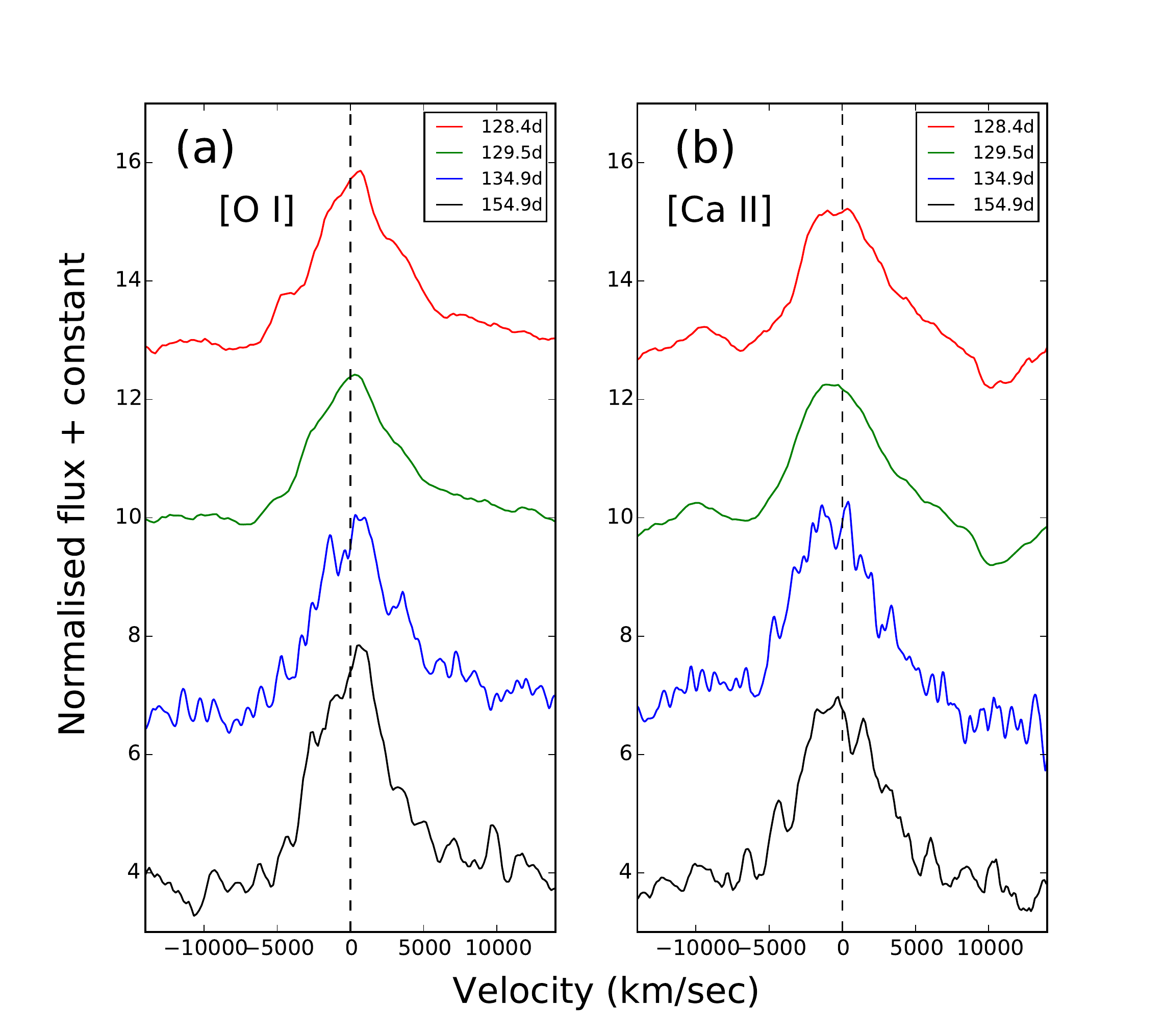}
	\end{center}
	\caption{Evolution of the nebular [O {\sc i}] and [Ca {\sc ii}] line profiles in SN 2015dj.}
	\label{fig:line_evolution_plot_SN 2015dj}
\end{figure}

One of the strongest emission features in type Ib SN nebular spectra is the [O {\sc i}] doublet at 6300 and 6363~\AA. This doublet provides an observational display of the explosion geometry along with other prominent features such as [Ca~{\sc ii}] and Mg~{\sc i}]. The [O~{\sc i}] feature is relatively isolated, whereas [Ca~{\sc ii}] and Mg~{\sc i}] lines are blended with other lines. Figure \ref{fig:line_evolution_plot_SN 2015dj} (panel a) illustrates the evolution of the [O~{\sc i}] line emission, whose peak shows a negligible shift from the rest wavelength in the velocity space: this is usually interpreted as an indication of spherically symmetric ejecta \citep{2009MNRAS.397..677T}.

Figure \ref{fig:line_evolution_plot_SN 2015dj} (panel b) presents the evolution of the [Ca~{\sc ii}] doublet 7291, 7324~\AA. It has a flat peak likely due to line blending. The strength of the [Ca~{\sc ii}] feature increases with time. Calcium clumps formed during explosion do not contribute considerably to the [Ca~{\sc ii}] emission \citep{1993ApJ...419..824L,2000AJ....120.1487M}. This feature is produced by the excitation of calcium present in the atmosphere by the pre-existing envelope. No evident shift from the rest velocity is observed for this emission feature, supporting spherical symmetry in the matter ejection (Figure \ref{fig:line_evolution_plot_SN 2015dj}, panel b)).

Following \cite{1986ApJ...310L..35U}, we calculate the mass of neutral oxygen which produces the [O {\sc i}] line, as 

\begin{equation}
	M_{O} = 10^{8} \times D^{2} \times F_{[O\,I]} \times e^{(2.28/T_{4})}
\end{equation}

\noindent                                            
where $M_{O}$ is the neutral oxygen mass in M$_{\odot}$ units, $D$ is the SN distance in Mpc, $F_{[O\,I]}$ is the oxygen line flux in erg~s$^{-1}$~cm$^{-2}$, and $T_{4}$ is the temperature of the oxygen emitting region, in units of 10$^{4}$~K. The above expression is valid in high density regime ($N_{e} \geq 10^{6}$ cm$^{-3}$) of type Ib SN ejecta \citep{1989AJ.....98..577S,1994AJ....108..195G,2004A&A...426..963E}. The temperature of this line-emitting region can be estimated by using the flux ratio of [O~{\sc i}] 5577/6300-6364 lines, which is 0.13 from the 155 day spectrum of SN 2015dj. The line-emitting region can have low temperature and high density ($T_4\leq$ 0.4) or high temperature and low density ($n_e\leq 5\times10^6$ cm$^{-3}$) for $T_4$ = 1 \citep{2007ApJ...666.1069M}. Since the [O~{\sc i}] emission occurs at high density and low temperature \citep{1989AJ.....98..577S,1991ApJ...372..531L,2004A&A...426..963E}, we assume $T_{4}$ = 0.4. The observed flux of the [O~{\sc i}] line doublet at 155 days is 2.10 $\times$ 10$^{-14}$~erg~s$^{-1}$~cm$^{-2}$; adopting a distance of 36.69~$\pm$~0.05~Mpc, we infer oxygen mass of 0.85~M$_{\odot}$.

We also find a weak O {\sc i} 7774 \AA{} in the nebular spectrum of SN 2015dj, which is indicative of residual ionized oxygen \citep{1986ApJ...302L..59B}.  The oxygen mass needed to produce the [O~{\sc i}] doublet and O~{\sc i}~7774\AA{} is higher than the mass of oxygen necessary to produce the [O~{\sc i}] doublet alone \citep{2010MNRAS.408...87M}. Therefore, our oxygen mass estimate should be treated as a lower limit for the total oxygen mass ejected in the explosion. The [O~{\sc i}] emission is produced because of a layer of oxygen formed during the hydrostatic burning phase. This oxygen mass is correlated with the main sequence progenitor mass \citep{1996ApJ...460..408T}. Following \citep{1996ApJ...460..408T} we thus estimate the progenitor and the He core masses to be 15--20~M$_{\odot}$ and 4--8~M$_{\odot}$, respectively.

Moreover, the flux ratio of [O~{\sc i}] and [Ca~{\sc ii}] lines can serve as a tool to probe the progenitor mass. \cite{2015A&A...579A..95K} investigated this ratio for core collapse SNe and found its insensitivity towards temperature and density. They also argued that the ratio is strongly dependent on the progenitor mass, and increases with larger progenitor mass \citep[see also][]{1989ApJ...343..323F,2004A&A...426..963E}. \cite{2015A&A...579A..95K} discussed the link of this ratio with different progenitor channels amongst type Ib/Ic SNe as they originate from massive WR stars or lower mass progenitors in binary systems. In type Ib/Ic SNe, this ratio is in the range $\sim$0.9 - 2.5, whereas in type II SNe it is $\textless$ 0.7 \citep{2015A&A...579A..95K}. In SN 2015dj, the [O~{\sc i}]/[Ca~{\sc ii}] flux ratio is between 0.72 and 0.86 inferred from the spectra at 128 and 155 days respectively. This flux ratio is in between the quoted values for type II and Ib/Ic SNe and indicates the association of SN 2015dj with a lower mass progenitor in a binary system.

\subsection{Progenitor mass from the nebular spectra modelling}
\label{Mass of progenitor from the nebular phase spectra modelling}

The mass of the progenitor strongly depends on the nucleosynthetic yields, which are in turn correlated with the strength of the nebular lines \citep{1995ApJS..101..181W}. Nucleosynthesis models for different progenitor masses were presented by \cite{2012A&A...546A..28J,  2014MNRAS.439.3694J}. \cite{2015A&A...573A..12J} incorporated some modifications in the code and extended the study to type IIb SNe. They included models for 12, 13 and 17 M$_{\odot}$. Since there is very little influence of the hydrogen envelope after $\sim$ 150 days, these models can be used as a comparison for type Ib SNe.   

\begin{figure}
	\begin{center}
		\includegraphics[width=\columnwidth]{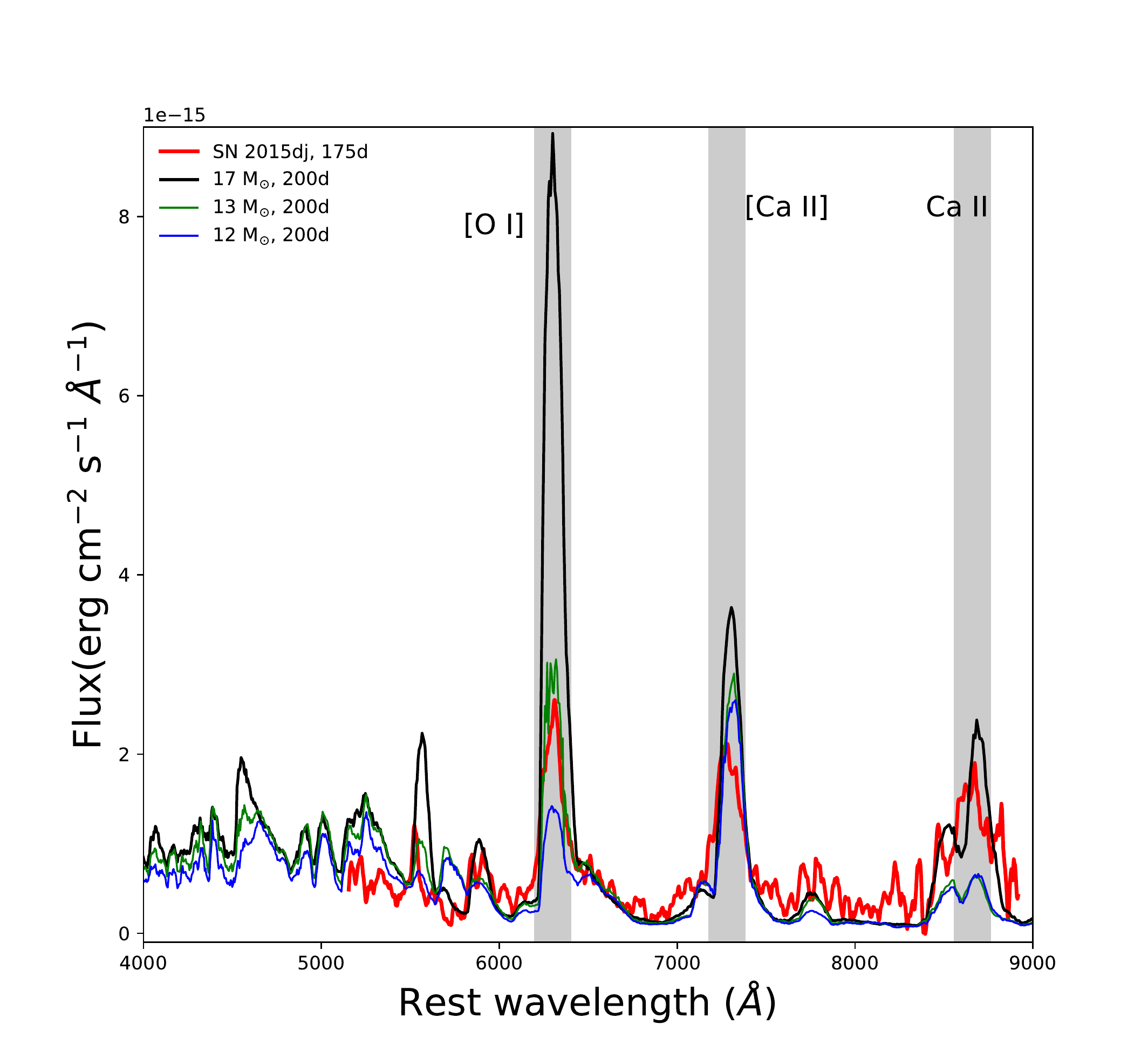}
	\end{center}
	\caption{Comparison of the late nebular spectrum of SN 2015dj with models of \citet{2015A&A...573A..12J} having different progenitor masses. Models have been scaled to match the properties of SN 2015dj in terms of [O~{\sc i}] luminosities. }
	\label{fig:SN 2015dj_spectral_fitting_jerkstrand_model.pdf}
\end{figure}

We compare the spectrum of SN 2015dj with the 200 days post explosion model spectra of \cite{2015A&A...573A..12J} having progenitor masses of 12, 13 and 17~M$_{\odot}$ (Figure \ref{fig:SN 2015dj_spectral_fitting_jerkstrand_model.pdf}). These models were produced for a $^{56}$Ni mass of 0.075~M$_{\odot}$ and a distance of 7.8~Mpc. The flux calibrated spectrum at {\bf $\sim$}175 day after explosion (155 day since V$_{max}$) of SN 2015dj is corrected for redshift and reddening. The model spectrum is scaled to match the $^{56}$Ni mass, distance and corresponding phase of SN 2015dj. We find that the luminosity of the [O~{\sc i}] feature of SN 2015dj is similar to 13 M$_{\odot}$ progenitor star. All three models underestimate the Ca~{\sc ii} NIR feature strength because of the assumption of a low density envelope.

The luminosity of the [O~{\sc i}] feature is a useful tool to probe the mass of nucleosynthesized oxygen. This is strongly dependent on the mass of the He core, hence the mass of the progenitor \citep{1995ApJS..101..181W,1996ApJ...460..408T}. Owing to the dominant role of the [O~{\sc i}] luminosity to estimate the progenitor mass, we expect it to be $\sim$ 13 M$_{\odot}$ (Figure \ref{fig:SN 2015dj_spectral_fitting_jerkstrand_model.pdf}). However, the lowest oxygen mass ejected in the explosion suggests that the progenitor mass should be in the range 15-20~M$_{\odot}$. These two measurements set a limit on the progenitor mass from 13 to 20~M$_{\odot}$ . 

\section{Summary}
\label{Summary}

We present an extensive photometric and spectroscopic evolution of a type Ib SN 2015dj. The light curve places this SN in the category of fast declining events among our comparison sample. The remarkable similarity of SN 2015dj with other type Ib SNe and late time spectral evidences rule out a type IIb SN classification. The light curve peak is well sampled in the {\it Vgri} bands, and the object reached a peak absolute magnitude of M$_V$ = $-$17.37~$\pm$~0.02 mag, and quasi-bolometric luminosity at maximum of 1.8~$\times$~10$^{42}$~erg~s$^{-1}$ which is consistent with other type Ib SNe. 

Analytical modelling of the quasi bolometric light curve of SN 2015dj gives $^{56}$Ni, M$_{ej}$ and E$_{k}$ of $0.06 \pm 0.01$ M$_{\odot}$, $1.4^{+1.3}_{-0.5}$ \msol\, and $0.7^{+0.6}_{-0.3} \times 10^{51}$ erg, respectively.

The spectral lines evolve faster in SN 2015dj than in similar events, and relatively low expansion velocities are found for the NIR [Ca~{\sc ii}] feature. The [O~{\sc i}] and [Ca~{\sc ii}] nebular lines have peaks centered at the rest wavelength, suggesting spherically symmetric ejecta. The nebular phase modeling and the inferred estimated [O~{\sc i}] mass indicate a progenitor mass of 13-20~M$_\odot$. The [O~{\sc i}] and [Ca~{\sc ii}] line ratio favours a binary scenario for SN~2015dj.

\section*{Acknowledgments}

We thank the referee for constructive comments on the manuscript which has improved the presentation of the paper.  We thank S. Taubenberger and P. Ochner for contributing towards observations of SN 2015dj. K.M. acknowledges the support from the Department of Science and Technology (DST), Govt. of India and Indo-US Science and Technology Forum (IUSSTF) for the WISTEMM fellowship and the Department of Physics, UC Davis, where a part of this work was carried out. KM also acknowledges BRICS grant DST/IMRCD/BRICS/Pilotcall/ProFCheap/2017(G) for the present work.  Research by SV is supported by NSF grants AST-1813176. NUTS is supported in part by the  Instrument  Center  for  Danish  Astrophysics (IDA). EC, NER, LT, SB and MT are partially supported by the PRIN-INAF 2016 with the project ``Towards the SKA and CTA era: discovery, localisation, and physics of transient sources" (P.I. M. Giroletti). We acknowledge Weizmann Interactive Supernova data REPository http://wiserep.weizmann.ac.il (WISeREP) \citep{2012PASP..124..668Y}. This research has made use of the CfA Supernova Archive, which is funded in part by the National Science Foundation through grant AST 0907903. This research has made use of the NASA/IPAC Extragalactic Database (NED) which is operated by the Jet Propulsion Laboratory, California Institute of Technology, under contract with the National Aeronautics and Space Administration. This work makes use of data obtained with the LCO Network, the Copernico telescope (Asiago, Italy) of the INAF - Osservatorio Astronomico di Padova and NOT.  CM, GH, and DAH were supported by NSF grant AST-1313484. 

\bibliography{ms}{}
\bibliographystyle{aasjournal}





\end{document}